\documentclass[12pt,preprint]{aastex}

\begin{document}

\title{The Origin of T Tauri X-ray Emission: New Insights from
the $Chandra$ Orion Ultradeep Project}

\author{Thomas Preibisch\altaffilmark{1},
Yong -Cheol Kim\altaffilmark{2},
Fabio Favata\altaffilmark{3},
Eric D. Feigelson\altaffilmark{4},
Ettore Flaccomio\altaffilmark{5},
Konstantin Getman\altaffilmark{4},
Giusi Micela\altaffilmark{5},
Salvatore Sciortino\altaffilmark{5},
Keivan Stassun\altaffilmark{6},
Beate Stelzer\altaffilmark{5,7},
Hans Zinnecker\altaffilmark{8}}

\altaffiltext{1}{Max-Planck-Institut f\"ur Radioastronomie,
 Auf dem H\"ugel 69, D-53121 Bonn, Germany}
\altaffiltext{2}{Astronomy Department, Yonsei University, Seoul, Korea}
\altaffiltext{3}{Astrophysics Division - Research and Science Support Department of ESA, ESTEC, Postbus 299, 2200 AG Noordwijk, The Netherlands}
\altaffiltext{4}{Department of Astronomy \& Astrophysics,
Pennsylvania State University, University Park PA 16802}
\altaffiltext{5}{INAF, Osservatorio Astronomico di Palermo G. S. Vaiana,
Piazza del Parlamento 1, I-90134 Palermo, Italy}
\altaffiltext{6}{Department of Physics and Astronomy, Vanderbilt University, Nashville, TN 37235}
\altaffiltext{7}{Dipartimento di Scienze Fisiche ed Astronomiche, Universit\`a
di Palermo, Piazza del Parlamento 1, I-90134 Palermo, Italy}
\altaffiltext{8}{Astrophysikalisches Institut Potsdam, An der Sternwarte 16, 
 D--14482 Potsdam, Germany}

\slugcomment{Version: \today}

\begin{abstract}
The $Chandra$ Orion Ultradeep Project (COUP) provides
the most comprehensive dataset ever acquired on the X-ray
emission of pre-main sequence stars. In this paper, we study 
the nearly 600 X-ray sources that can be reliably identified with 
optically well characterized T Tauri stars (TTS) in the
Orion Nebula Cluster. With a detection limit of 
$L_{\rm X,min} \sim 10^{27.3}$~erg/sec for lightly absorbed 
sources, we detect X-ray emission from more than 97\% of the 
optically visible late-type (spectral types F to M) cluster stars. 
This proofs that there is no ``X-ray quiet'' population of 
late-type stars with suppressed magnetic activity.
We use this exceptional optical, infrared, and X-ray data set
to study the dependencies of the X-ray properties on other
stellar parameters. 
All TTS with known rotation periods lie in the saturated 
or super-saturated regime of the relation between activity and 
Rossby numbers 
seen for main-sequence (MS) stars, but the TTS show a much larger 
scatter in X-ray activity than seen for the MS stars.
Strong near-linear relations between X-ray luminosities, bolometric
luminosities and mass are present.
We also find that the fractional X-ray
luminosity $L_{\rm X}/L_{\rm bol}$ rises slowly with mass over the
$0.1 - 2\,M_\odot$ range.
The plasma temperatures determined from the X-ray spectra of the 
TTS are much hotter than in MS stars, but seem to follow a general
solar-stellar correlation between plasma temperature and 
activity level. The scatter about the relations between 
X-ray activity and stellar parameters
is larger than the expected effects of X-ray variability, 
uncertainties in the variables, and unresolved binaries.
This large scatter seems to be related to the
influence of accretion on the X-ray emission. 
While the X-ray activity of the non-accreting  TTS is consistent 
with that of rapidly rotating MS stars,
the accreting stars are less X-ray active (by a factor of 
$\sim 2-3$ on average) 
and produce much less well
defined correlations than the non-accretors.
We discuss possible reasons for the suppression of X-ray 
emission by accretion
and the implications of our findings on 
long-standing questions related to the origin of the
X-ray emission from young stars, considering in particular 
the location of the X-ray emitting structures and 
inferences for pre-main-sequence magnetic dynamos.

\end{abstract}


\section{Introduction \label{intro.sec}}
\subsection{X-ray emission from young stellar objects}

Young stellar objects (YSOs) in all evolutionary stages from 
class~I protostars 
to ZAMS stars show highly elevated levels of X-ray 
activity (for recent reviews on the X-ray properties of YSOs
and on stellar coronal astronomy in general see Feigelson \& Montmerle 1999 and
Favata \& Micela 2003).
X-ray observations of star forming regions 
allow to study the high energy processes in YSOs,
which are of great importance for our understanding of
the star formation process. For example, the X-ray emission from a
YSO should
photoionize its circumstellar material and thus influence accretion
as well as outflow processes, both of which are thought
to be based on the interaction of ionized material with magnetic
fields. The X-ray emission from the central YSO is certainly an important,
probably even the dominant factor in determining the ionization structure of
protoplanetary disks, and has therefore a strong impact on processes
like the
formation of proto-planets \citep[e.g.][]{Glassgold00, Matsumura03}.
 
The first discoveries of X-ray emission from T~Tauri~stars (TTS, $=$
low-mass pre-main sequence (PMS) stars) 
were made with the EINSTEIN X-ray observatory 
\citep[e.g.][]{Feigelson81} and revealed a surprisingly 
strong X-ray activity, exceeding the solar levels by several orders of 
magnitude. The X-ray observations also revealed a new population of 
young stellar objects \citep{Walter88}, the ``weak-line T Tauri stars'' (WTTS),
which lack the classical optical signposts of youth, like strong 
H$\alpha$ emission, of the previously known ``classical T Tauri stars'' (CTTS).
The ROSAT observatory 
increased the number of observed star forming regions, and thereby the
number of known X-ray emitting TTS, considerably
\citep[e.g.][]{Feigelson93, Casanova95, Gagne95, Neuhaeuser95, Preibisch96}.
The ROSAT All Sky Survey lead to the 
X-ray detection
of extended populations of TTS in and around many star forming
regions \citep[e.g.][]{Neuhaeuser97} and
 demonstrated that the stellar populations
of star forming regions are considerably larger than suspected by earlier
surveys based on classical youth indicators such as H$\alpha$ emission.
The ROSAT All Sky Survey was also well suited to study the X-ray properties
in complete, volume limited samples of nearby field stars. An important result from
such studies was
that apparently all cool dwarf stars are surrounded by X-ray emitting coronae,
with a minimum X-ray surface flux around $10^4$~erg/sec/cm$^2$ 
\cite{Schmitt97}.
The ASCA satellite 
detected X-ray emission from numerous deeply embedded YSOs;
due to its rather poor spatial resolution, however, the proper identification
of the X-ray sources was often difficult.

While the X-ray missions of the last two decades provided important information
about the X-ray properties of YSOs, there were also serious limitations.
First,
the typical samples of X-ray detected objects in young clusters and
star forming regions
contained hardly more than $\sim 100$ objects, too few to allow
well founded statistical conclusions to be drawn.
Second, a large fraction of the known cluster members (especially 
low-mass stars) remained undetected 
in X-rays, and any correlation studies had therefore to deal with 
large numbers of upper limits.
Third, especially in dense clusters, the individual sources could often not be 
spatially resolved, and so the proper identification of the X-ray sources was 
difficult or impossible.
Finally, only a relatively small number of individual young stars
were bright enough in X-rays to allow their spectral and temporal X-ray 
properties to be studied in detail, and it is not clear whether these stars 
really are ``typical'' cases or perhaps peculiar objects.

With the advent of $Chandra$ and {\it XMM-Newton}, 
the situation has improved substantially.
Due to the large collecting areas of these observatories, 
their sensitivity is at least an order of magnitude better than that of
earlier missions.
Due to their wide energy band, extending from $\sim 0.2-0.5$~keV 
up to  $\sim 8-10$~keV, they are very well suited to study the 
hard X-ray emission from highly obscured  YSOs \citep[e.g.,][]{Skinner03}.
 Furthermore, $Chandra$ has a superb point spread function, 
providing a spatial resolution of better than $1''$;
this abolishes the usual identification problems in nearby
star forming regions.

\subsection{Open questions about the X-ray activity from TTS}

We enunciate two basic, still unresolved questions concerning
the origin of the elevated X-ray activity of TTS:
Does the strong X-ray activity of TTS, with
X-ray luminosities up to $\sim 10^4$ times and 
plasma temperatures up to $\sim 50$ times higher than seen in our Sun,
originate from solar-like coronae? If so,  are these coronae created
and heated by
solar-like (although strongly enhanced) magnetic dynamo processes,
or are fundamentally different magnetic structures 
and heating mechanisms involved?

One main obstacle on the way towards an understanding of TTS X-ray activity
is the fact that the solar corona has an extremely complex and dynamic
structure with many different facets \citep[e.g.,][]{Aschwanden01}; 
it is not clear to what
degree comparisons and extrapolations from the solar to the stellar case
make sense. A second problem is that even for the Sun,
which can be studied in great detail
at high spatial, temporal, and spectral resolution, the important
question about the heating of the solar corona remains puzzling,
even after 5 decades of intense research \citep[e.g.,][]{Walsh03}.
The third problem is our lack of a sound understanding of the dynamo
processes which are the ultimate origin of the magnetic activity in the
Sun and in stars \citep[e.g.,][]{Ossendrijver03}.

The X-ray activity
of main-sequence (MS) stars is mainly determined by
their rotation rate. The well established rotation--activity relation
\citep[e.g.][]{Pallavicini81, Pizzolato03}
is given 
by the power-law relation $L_{\rm X}/L_{\rm bol} \propto P_{\rm rot}^{-2.6}$,
in agreement with the expectations from
 solar-like $\alpha\!-\!\Omega$ dynamo models \citep[e.g.][]{Maggio87}.
At periods shorter than $\sim$~2-3 days, the activity saturates at
 $\log\left(L_{\rm X}/L_{\rm bol}\right) \sim -3$ for reasons that
are not yet understood.
The plasma temperatures generally increase with the level of X-ray
activity, scaling roughly as
$T_{\rm X} \propto \left(L_{\rm X}/L_{\rm bol}\right)^{0.5}$ \citep[e.g.][]{Preibisch97}.
Most TTS rotate quite rapidly, and neither their X-ray luminosities nor
their plasma temperatures are unusual when compared to rapidly rotating
MS stars.

However, a relation between rotation and X-ray activity could never be
convincingly established for TTS; in most studies the small number of
X-ray detected TTS with known rotation periods did not allow to draw
sound conclusions. This problem of 
small and often biased samples has, however, recently been overcome
with two $Chandra$ studies of the ONC \citep{Feigelson02a, Flaccomio03b}, 
both of which found {\em no significant relation between 
X-ray activity and rotation}.
This strongly puts into question the solar-like
dynamo activity scenario for TTS.
Another argument against solar-like dynamos comes from theoretical
considerations: at ages of only a few Myr, the TTS
are usually thought to be  fully convective, and therefore the
standard solar-like $\alpha\!-\!\Omega$ dynamo, which is anchored
at the boundary between the convective envelope and the inner radiative
core, should not work. 
Theoreticians have developed alternative dynamo concepts 
(e.g., K\"uker \& R\"udiger 1999; Giampapa et al.~1996) 
that may work in fully convective stars. A problem with these and other
models is that they usually do not make quantitative predictions
that can be easily tested from observations.
Further possibilities for the origin of X-ray emission from TTS 
include magnetic fields coupling the stars to their surrounding 
circumstellar disk (see, e.g., Hayashi et al.~1996; Montmerle et al.~2000;
\citealp{Iobe03}; \citealp{Romanova04}), 
or X-ray emission from accretion shocks
\citep[see, e.g.,][]{Kastner02, StelzerSchmitt04, Favata03, Favata05}.
The investigation of these possibilities in the light of the
\dataset[ADS/Sa.CXO#obs/COUP]{the {\it Chandra\/} Orion
Ultradeep Project (COUP)} data
will be a major topic of this study.

\subsection{Properties of the ONC and previous X-ray observations}

The Orion Nebula is an HII region on the near side of a giant molecular
cloud, which contains one of the most prominent and nearby ($D \sim 450$ pc) 
star forming regions (for a recent review see \citealp{Odell01}). 
This star forming region contains a massive  cluster of young
($\approx 10^6$ yr) stars \citep[cf.][]{Herbig86, McCaughrean94, Hillenbrand97}, which is known as the
Orion Nebula Cluster (ONC).  
The Orion Nebula is illuminated mainly by the two 
O-type stars $\theta^1{\rm Ori}$\,C and $\theta^2\,{\rm Ori}$\,A.
Since the ONC is a perfect laboratory for observations of
 star formation over the full stellar mass range,
it is one of the best investigated star forming regions
and has been observed at virtually any wavelength.
\cite{Hillenbrand97} has compiled a catalog of nearly 1600~optically visible stars
within $\sim 2.5$~pc of the Trapezium; for over 900 of these  stars
enough information is avaliable to place them into the HR-diagram
and to determine their masses and ages by comparison with theoretical
PMS evolution models.

The ONC has been observed with basically all previous X-ray observatories
(see e.g.~\cite{Ku79} for EINSTEIN observations; 
 \cite{Gagne95, Geier95, Alcala96} for ROSAT observations;
\cite{Yamauchi96} for ASCA observations).
However, the high spatial density of stars in the ONC 
and the poor spatial resolution of these X-ray observatories did not allow
a reliable identification of many X-ray sources.
 Only $Chandra$ with its
superb point spread function is suitable for studying the ONC, where
the mean separation between the sources in the inner $1'$ radius area is 
only $5''$.
The ONC has been observed with both imaging instruments onboard of $Chandra$.
The results of two ACIS-I observations with a combined 
exposure time of 23 hours were reported in 
 \cite{Garmire00} and \cite{Feigelson02a, Feigelson02b, Feigelson03}. 
1075 individual sources were detected, 91\% of which 
could be identified with known stellar members of the cluster.
\cite{Flaccomio03a, Flaccomio03b}
presented the analysis of a 17.5 hr HRC-I
observation of the ONC, which yielded 742 X-ray sources in the
$30' \times 30'$ field-of-view.
Furthermore, some of the brightest X-ray sources in the ONC have also
been studied with the High Energy Transmission Grating Spectrometer
\citep{Schulz00, Schulz01, Schulz03}, but most of these sources are massive stars which are
not the topic of this paper.

\bigskip

In this paper, we discuss the X-ray data on the TTS in the ONC resulting
from COUP, by far the longest and most sensitive X-ray observation
ever obtained for the ONC.
The plan of this paper is as follows: after briefly describing the COUP 
observation in \S\ref{obs.sec}, we define in \S\ref{opt_sample.sec}
the optical sample which will be the basis of our studies,
and then investigate the relation of the X-ray emission to 
basic stellar parameters in \S\ref{relations.sec}.
In \S\ref{rotation.sec} we study in detail the relation between
X-ray emission, rotation and convection.
In \S\ref{variability.sec} we discuss the origin of the large scatter
seen in the correlations  between X-ray activity and other stellar parameters.
In \S\ref{spectra.sec}, we investigate the plasma temperatures
as determined from the fits to the X-ray spectra.
Section~\ref{accretion.sec} deals with the possible connections between
 X-ray emission and circumstellar accretion disks. 
Finally, in \S\ref{conclusions.sec} we discuss the implications of
our results with respect to the origin of the TTS X-ray emission.

%%%%%%%%%%%%%%%%%%%%%%%%%%%%%%%%%%%%%%%%%%%%%%%%%%%%%%%%%%%%%%%%%%%%%%

\section{The COUP observation\label{obs.sec}}

The COUP observation is the deepest and longest X-ray observation ever made
of a young stellar cluster, providing a rich and unique 
dataset for a wide range of science studies.  
The observational details and a complete description of the 
data analysis can be found in \citet{Getman05a}, here we summarize
only the aspects that are most important to our studies.
The COUP observation was performed between 8 Jan 2003 and 21 Jan 2003,
utilizing the Advanced CCD Imaging Spectrometer (ACIS) in its imaging 
configuration, which gives
a field of view of  $17' \times 17'$. 
The total exposure time of the COUP image was 838\,100 sec (232.8 hours
or 9.7 days).
The spatial resolution of ACIS is better than $1''$ over most of the
field of view. 
The very low background allows the reliable detection 
of sources with as little as $\sim 5$ source counts.
The final COUP source catalog lists 1616 individual sources.
The superb point spread function and the high accuracy of the aspect solution allowed a
clear and unambiguous identification of nearly all X-ray sources 
with optical or near-infrared counterparts\footnote{The median offsets 
between COUP sources and near-infrared or optical
counterparts are only 0.15\arcsec and 0.24\arcsec, respectively}.

Spectral analysis was performed using a semi-automated approach to produce
an acceptable spectral model for as many as possible sources.
The XSPEC spectral fitting programme was used to fit the extracted spectra
with one- or two-temperature optically thin thermal plasma MEKAL models
assuming 0.3 times solar abundances and X-ray absorption.
The parameters derived in these fits are the hydrogen column density
$N_{\rm H}$ as a measure of the X-ray absorption, and the temperatures
$T_{\rm X}$ and emission measures $EM$ of the one or two spectral
components.
The spectral fitting results were also used to compute the 
intrinsic (extinction-corrected) X-ray luminosity by integrating the model
source flux over the $0.5\!-\!8$~keV band. 

Our analysis in this paper is based on the tabulated X-ray properties 
and identifications of the COUP sources as listed in \citet{Getman05a}. 
We use the identifications of the X-ray sources 
with optical counterparts as given in their Table~9, the X-ray luminosities 
and X-ray spectral properties as listed in their Tables~8 and 6, 
and upper limits of undetected stars in the Hillenbrand sample as given in 
Table~11.

The temporal behavior of COUP sources is often very complex, with
high-amplitude, rapidly changing flares superposed on apparent quiescent or
slowly variable emission as studied in detail by Wolk
et al.~(2005), Favata et al.~(2005) and Flaccomio et al.~(2005).
For the purpose of the present study, the individual features of the 
lightcurves are not of interest.
We note that the X-ray properties tabulated in the COUP tables,
i.e.~the count rates, derived spectral parameters, and X-ray luminosities,
represent the average over the 10 days exposure time of our dataset.
This implies that the effect of short excursions in the X-ray lightcurves,
like flares with typical timescales of a few hours, are strongly 
``smoothed out''.
Since most of the young stars are rather fast rotators with
periods of less than 10 days, the tabulated X-ray properties represent 
for these objects the
average over at least one rotation period and therefore also smooth out
possible rotational modulation.

Nevertheless, it would be interesting to establish
the level of ``quiescent'' X-ray emission in the sources,
i.e.~the sustained, or ``typical'' level of X-ray emission outside the periods
of flares or otherwise elevated activity.
For this purpose, we used the results of the
Maximum Likelihood Blocks (MLB) lightcurve 
analysis \citep{Flaccomio05}, which segments the
lightcurves into contiguous sequences of constant count rates and allows
to discern between periods of flaring and more constant, sustained
X-ray emission. 
A strict and fully convincing definition for ``quiescent'' X-ray emission 
is not possible, especially since any apparently quiescent emission
may, in reality, just be a superposition of numerous unresolved small flares.
Wolk et al. (2005)
empirically establish a proxy for the quiescent emission levels
by determining a "characteristic level"
in each lightcurve, which is essentially defined
as the average count rate over periods where the count rate is not significantly
elevated.
An estimate for the characteristic X-ray luminosity
can then be obtained by multiplying the temporally averaged X-ray luminosity,
as determined from the spectral analysis,
with the ratio of the characteristic countrate from the MLB analysis to the
mean countrate over the COUP exposure.
We note that this simple scaling procedure 
is not fully self-consistent, because it does not take into account
that the X-ray spectral
parameters (and thereby the transformation factor from count rate to
luminosity) can change as a function of the emission level, but it
should be appropriate for our purposes.

The difference between the average and characteristic
X-ray luminosities is generally not large:
the median value of the correction factor is 0.78;
only for 13\% of the TTS this factor is
$<1/2$, and for only 4\% of the TTS $< 1/3$.
We will show below that the choice of either the average or the 
characteristic
X-ray luminosities has generally very little effect on the observed relations.
We will therefore mainly use the temporally averaged X-ray luminosities
and consider the characteristics luminosities only in a few
cases.

\section{Definition of the optical sample and X-ray detection completeness\label{opt_sample.sec}} 

\subsection{The optical sample of ONC stars}

The aim of our study is to investigate the X-ray properties of
a homogenous and well defined sample of comprehensively characterized 
TTS (= young late type [F--M] stars).
We will therefore {\em not} consider the COUP detected 
 brown dwarfs \citep[see][]{Preibisch05}
or embedded objects \citep[see][]{Grosso05}, or 
OBA stars \citep[see][]{Stelzer05},
although we will sometimes compare the more massive stars with TTS.

The basis for the construction of our ``optical sample'' is the 
\cite{Hillenbrand97} [H97 hereafter] 
sample of $1576$ optically visible ($I \lesssim 17.5$) stars within $\sim 2.5$~pc
($\sim 20'$)
of the Trapezium, for $934$ of which optical spectral types are known. 
We used an updated version of the H97 tables
in which for many objects spectral types and other stellar parameters
have been revised\footnote{\url{http://www.astro.caltech.edu/\~\,lah/papers/orion\_main.table1.working}}.
Some of the stars in this area are unrelated field stars
lying either in the foreground or the background of the Orion Nebula; these
should of course be excluded from our studies.
We therefore used the membership information from proper motion studies 
listed in the table of H97: we consider all stars with membership 
probabilities $\ge 50\%$ to be bona-fide members of the ONC, 
while stars with membership probabilities $< 50\%$ are
considered here to be non-members and excluded from
our analysis.
Stars with no membership information are considered
here as likely cluster members, because contamination by
foreground field stars is very small\footnote{H97
estimate that $\sim 97\%$ of the $I\lesssim 17$ mag stars within
about 1 pc of the Trapezium (i.e., roughly the field of view of our COUP image) 
are ONC members},
and contamination by background stars
is unlikely due to the large visual extinction 
in the molecular could immediately behind the ONC.
H97 assumes that the optical database is {\em representative of all stars
in the ONC region} and that the completeness of $\sim 60\%$ is
 uniform with radius.
 
1056 of the H97 stars are within the field-of-view of the COUP observation,
892 of which are detected as X-ray sources. Excluding the 33 stars
that are identified as non-members, we have 1023 likely ONC members 
from the H97 sample in the COUP field-of-view and detect 870 of these
(i.e., 85\%) as X-ray sources.
For the analysis in this paper we use those of these stars 
for which spectral types are known.
Our ``optical sample'' then consists of  
639 optically visible likely ONC members from H97 with known spectral type;
598 of these stars (i.e., 94\%) are detected as X-ray sources,
 41 remain undetected in the COUP image. 
The spectral types range from O7 for $\theta^1\,C$~Ori,
the most massive and luminous star in the ONC, down to 
$\sim$ M6.5 for objects close to the stellar-substellar 
boundary (at $\sim 0.075\,M_\odot$) at the age of the ONC.
For 575 stars in the COUP optical sample bolometric 
luminosities are known, allowing them to be placed into the HR-diagram.
 Masses and ages were estimated for 536 stars
by comparison of their location in the HR-diagram
to the theoretical PMS evolutionary tracks from \cite{Siess00}.
The masses in the COUP optical sample range from $0.1\,M_\odot$
(the lowest mass in the \cite{Siess00} models)
to $38\,M_\odot$ for $\theta^1$~C~Ori.
 
The visual extinction is known for 631 of the 639 
stars in our optical sample and varies from
0 to $A_V = 11$~mag,
with a mean value of $\langle A_V\rangle = 1.55$~mag.
Since the optical sample is magnitude limited, extinction
introduces a bias, since intrinsically brighter stars can suffer from
more extinction and still be included in the sample than
the intrinsically fainter stars.
We therefore used an extinction limit to construct a more homogeneous
sample:
we define as the ``lightly absorbed optical sample'' those stars
for which the optical extinction is known and is
$A_V \le 5$~mag. The lightly absorbed optical sample consists
of 586 stars, 554 of which are detected as X-ray sources in the
COUP image.
This extinction limit also yields a rather uniform sample with respect
to the X-ray detection limit: PIMMS simulations for Raymond-Smith spectra
with $kT = 2$~keV show that the detection limit (i.e.~the X-ray luminosity
that corresponds to a given number of detected source counts) increases 
by $0.39$~dex when going from zero extinction to $N_{\rm H} = 8 \times
10^{21}\,\rm cm^{-2}$ ($A_V = 5$~mag).
Since the uncertainty of the X-ray luminosity determinations 
is similar to this factor, our extinction limited
sample does not suffer from a significant 
extinction-dependent X-ray detection bias.

To summarize, our ``lightly absorbed optical sample'' of 586 stars is not
100\% complete (because spectral types are not available for all stars
in the ONC), but nevertheless should be a {\em statistically representative
sample of the ONC young stellar population with low extinction}.
The only potential systematic
selection effect might be that older ($\gtrsim 10$~Myr) 
very-low mass
($M \lesssim 0.2\,M_\odot$) stars may be missing; it is, however, unclear
whether such an older population of ONC members does exist at all.

For the 42 stars in the optical sample which were not detected as
X-ray sources in the COUP data, we estimated upper limits to their
X-ray luminosities from the tabulated upper limits 
to their count rates following the procedure outlined
in \citet{Getman05a}.

\subsection{X-ray detection completeness}

Table~\ref{detfrac.tab} lists the COUP X-ray detection fractions for the different
spectral types. It is important to note here that most of the non-detections
of ONC stars are due to X-ray source confusion in the COUP data; the typical
case are close ($\sim 1''-2''$ separation) binary systems, in which only one of the
components is clearly detected as an X-ray source \citep{Getman05a}.
In these cases the object would perhaps have been detected if located
at a different position. Since the occurrence of source confusion should not
depend on stellar parameters, these objects 
can be considered as ``unobserved'' and will be ignored in our analysis. 
We can thus compute an ``effective'', confusion-free
detection fraction by removing these non-detections with source confusion from
the sample. Then, the effective detection fractions  range between
97\% and 100\% for all spectral classes except the A- and B-type stars.
 This means that less than 3\% of the TTS in the optical sample
are undetected because their X-ray emission is below our detection 
limit.

With so few undetected objects, we can look at these stars in detail.
The undetected B8 star H97-1892 and the A1 star H97-531 
are intermediate mass stars and will be discussed in  
Stelzer et al.~(2005).

The three undetected K-type stars not suffering from source confusion
are H97-62, H97-489, and H97-9320.
H97-62 has a rather large extinction of $A_V = 5.26$~mag that may have
absorbed too much of its X-ray emission.
H97-489 has no bolometric luminosity and optical extinction
listed in H97.
The third object is
the K6-star H97-9320, which lies far (2.7 mag) below the ZAMS in the 
HR-diagram,
putting considerable doubt on its membership to the ONC;
since the star has also no proper-motion membership information in H97,
we suspect it to be a non-member and exclude it from our optical sample.
We therefore conclude that {\em all K-type stars in the lightly 
absorbed optical sample are detected}.

For the 13 undetected M-type stars not suffering from source confusion
we note that two objects have optical extinction
exceeding $A_V = 5$~mag, and 
4 objects have no proper-motion membership information in H97 and
may therefore perhaps be non-members. 
Only 8 of the M-type stars among the known proper motion 
members in the lightly absorbed optical
sample remain undetected.
The upper limits to the fractional X-ray luminosities of the 
undetected M-type stars range from 
$\log\left(L_{\rm X}/L_{\rm bol}\right) < -5.46$ (for H97-305) to 
$\log\left(L_{\rm X}/L_{\rm bol}\right) < -4.37$ (for H97-853).
This is considerably below the mean fractional X-ray luminosities 
of the detected M-type stars of 
$\log\left(L_{\rm X}/L_{\rm bol}\right) = -3.62$,
but still within the range of fractional X-ray luminosities
found for the detected M-type stars, three of which have
values below the lowest upper limit for the non-detections.
We conclude that the very few undetected M-type stars show low, 
but not necessarily unusually low levels of X-ray activity.

To summarize, we find 
that COUP detects every optically visible star
in the ONC sample except a few of the intermediate mass stars
(which are not expected to be intrinsic X-ray emitters)
and a few of the M-type stars (some of which may perhaps be non-members).
{\em We find no indications for the existence of an ``X-ray quiet''
population of stars with suppressed magnetic activity.}
Our analysis is thus based on a (nearly) complete sample, and
we can be very confident that our conclusions will not be affected by
non-detections. The only remaining concern is about the completeness
of the optical sample, which may not be very well established
for older ($\gtrsim 10^7$~yr) very low-mass ($\lesssim 0.2\,M_\odot$) stars.

\bigskip

It is interesting to note that most (1047 of 1616,
i.e.~65\%) of the
COUP X-ray sources were already detected in the previous 23 hr exposure
ACIS observation or the 17.5 hr HRC observation of the ONC.
Since the field-of-view
covered by the different observations is not identical, we focus on the
inner $8'$ radius area, which is included in all $Chandra$ observations
considered here.
In this area, % inner $8'$ radius field-of-view,
970 (66.6\%) of the 1457 COUP sources,
475 (90.9\%) of the 522 COUP sources in the optical sample, and
438 (91.3\%) of the 480 COUP sources in the lightly absorbed optical sample
were already detected in the previous
23~hr ACIS observation or the 17.5~hr HRC observation.
With more than 10 times the earlier exposure times,
COUP leads to the detection of
only a relatively small number of 47 (42) new X-ray sources that could be 
identified with stars in the
(lightly absorbed) optical sample. 
The COUP data nevertheless represent an important step forward over the
previous observations, since they 
increased the X-ray detection fraction in the lightly absorbed
optical sample from $\sim 88\%$ to at least 97\%, transforming an
incomplete sample to a (nearly) complete sample. Also,
the COUP data for the individual sources have much higher S/N 
(counts per source) than previous data sets,
thus allowing a much more reliable determination of the X-ray source
properties.

\section{Relation of the X-ray emission to basic stellar parameters\label{relations.sec}}

Since the origin of the X-ray activity in TTS is still not well known,
it is unclear which are the best parameters to consider in
making correlations.
We therefore consider several possibly useful stellar parameters 
(bolometric luminosity, stellar mass, effective temperature, rotation,
circumstellar disk properties, accretion rates)
to look for relations to the X-ray emission level;
note that relation between X-ray activity and age are discussed in a separate
paper (Preibisch \& Feigelson~2005).
For the characterization of the X-ray properties we consider here
the X-ray luminosity $L_{\rm X}$,
the fractional X-ray luminosity $L_{\rm X}/L_{\rm bol}$, 
and the X-ray surface flux $F_{\rm X}$, i.e.~X-ray luminosity divided
by the stellar surface area.
Throughout this paper, we use
$L_{\rm X}$ to refer to the extinction-corrected total band ($0.5-
8$~keV) luminosity $L_{t,c}$ defined and listed by \citet{Getman05a}.

Most of the relations presented here were already studied in other
X-ray data sets \citep[e.g.][]{Feigelson03, Flaccomio03b, PZ02}, 
often with similar results to what we find here.
Nevertheless, we study these relations here in some detail because
our COUP data provide a unique, in fact the best data set 
for an investigation of the
nature and the origin of the X-ray emission from TTS
for the following reasons.
With the {\em exceptionally well characterized young stellar population}
in the ONC
we can take advantage of known stellar parameters  for several hundred stars.
The {\em high sensitivity of the COUP X-ray data} yields a
 detection limit of
$L_{\rm X,min} \sim 10^{27.3}$~erg/sec for lightly absorbed stars and
 allows us to detect more than 97\% of the stars in the
lightly absorbed optical sample of cluster members.
Our analysis is therefore based on an {\em nearly
complete sample}. The high sensitivity allows us to detect X-ray emission
of the young solar-luminosity stars down to activity 
levels\footnote{Note that the quoted activity level
refers to the relatively hard 0.5--8~keV COUP band. This band covers
most of the X-ray flux from TTS (which are characterized by rather high 
plasma temperatures of $\ga 10$~MK and have accordingly relatively hard 
X-ray spectra). Solar-like field stars and the Sun, however, exhibit
considerably lower plasma temperatures ($\sim 2$~MK in the case of the Sun)
and have accordingly softer X-ray spectra with most of the X-ray flux 
below the COUP band.
If the Sun were located at the distance of the ONC, it would be only
marginally detectable during its maximum phase of coronal activity.
 }  of
$\log\left(L_{\rm X}/L_{\rm bol}\right) \leq -6$.
Our statistical analysis strongly benefits from the
{\em large sample} of 598 optically well characterized ONC stars
for which X-ray emission is detected.
This represents the largest homogenous sample of TTS that
has ever been studied
with very sensitive X-ray observations (and will remain so for the
foreseeable future).
The {\em availability of X-ray spectra} with good S/N for
nearly all sources allows the determination of reliable X-ray luminosities.
The {\em 10 day long observation}
provides a much better measure for the
``typical" X-ray properties of the strongly variable TTS
than observations with shorter exposure times, which yield only
a ``snapshot".

\subsection{X-ray luminosity and bolometric luminosity}

\begin{figure}[p]
\plotone{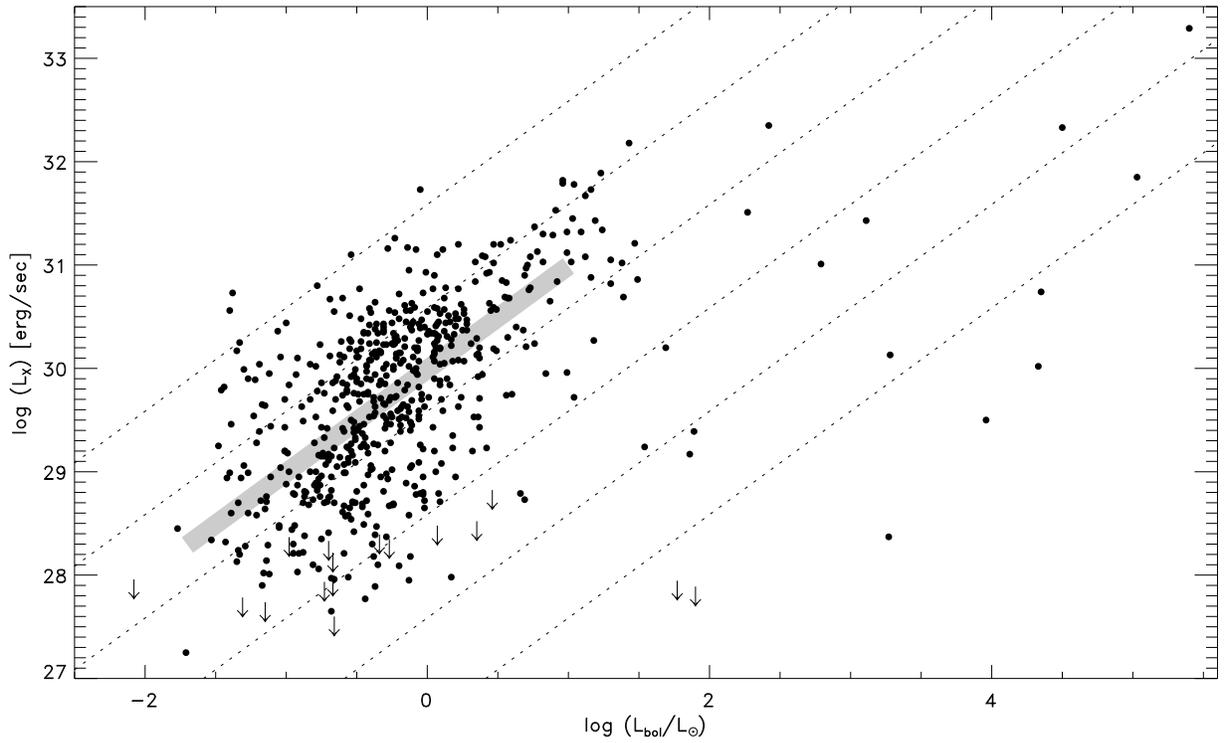}
\figcaption{X-ray luminosity versus bolometric luminosity for the stars
in the optical sample. For the members of the optical sample which are
not detected as X-ray sources in the COUP data the arrows show the upper
limits to their X-ray luminosities.
The dotted lines mark $\log\left(L_{\rm X}/L_{\rm bol}\right)$ ratios of $-2$,
$-3$, $-4$, $-5$, $-6$, and $-7$.
The thick grey line shows
the EM
algorithm linear regression fit for the
$L_{\rm X}\leftrightarrow L_{\rm bol}$ relation for
$L_{\rm bol} \leq 10\,L_\odot$ stars computed with ASURV.
\label{lx_lbol.fig}}
\end{figure}

The plot of X-ray luminosity versus 
bolometric luminosity is shown in Fig.~\ref{lx_lbol.fig}.
Nearly all stars  show
$\log\left(L_{\rm X}/L_{\rm bol}\right) > -5$ and therefore are much more
X-ray active than the Sun (for which $\log\left(L_{\rm X}/L_{\rm bol}\right) 
\sim -6.5$
is an average during the course of the solar cycle). 
The most active stars show fractional X-ray luminosities around
$\log\left(L_{\rm X}/L_{\rm bol}\right) = -3$, which is known as the ``saturation limit'' 
for coronally active stars \citep{Fleming95}.

Considering only the low-luminosity ($L_{\rm bol} < 10\,L_\odot$) stars,
we find a clear correlation between X-ray and bolometric luminosity,
although with a very large scatter.
We utilized the ASURV survival analysis package 
\citep{Feigelson85, Iobe86, LaValley90}
for the statistical investigation of the relation between $L_{\rm X}$
and $L_{\rm bol}$.
The ASURV software allows one to deal with data sets that contain 
non-detections 
(upper limits) as well as detections, and provides the maximum-likelihood
estimator of the censored distribution,
several two-sample tests, correlation tests and linear regressions.
The linear regression fit with the parametric Estimation Maximization (EM)
 algorithm in ASURV  yields
$\log\left(L_{\rm X}\,[{\rm erg/sec}]\right) = 30.00(\pm 0.04) + 1.04(\pm0.06)\times\log 
\left(L_{\rm bol}/L_\odot\right)$ with a standard deviation of 0.70~dex
in $\log L_{\rm X}$ for the low-luminosity stars
($L_{\rm bol} \leq 10\,L_\odot$).
This relation 
is very similar to the relations
found for other young clusters (cf.~Feigelson \& Montmerle 1999)
and is consistent with a linear relation between X-ray and bolometric
luminosity characterized by $\langle \log\left(L_{\rm X}/L_{\rm bol}\right) 
\rangle = -3.6 \pm 0.7$.

\subsection{X-ray activity and stellar mass\label{lx_mass.sec}}

\begin{figure}[p]
\plotone{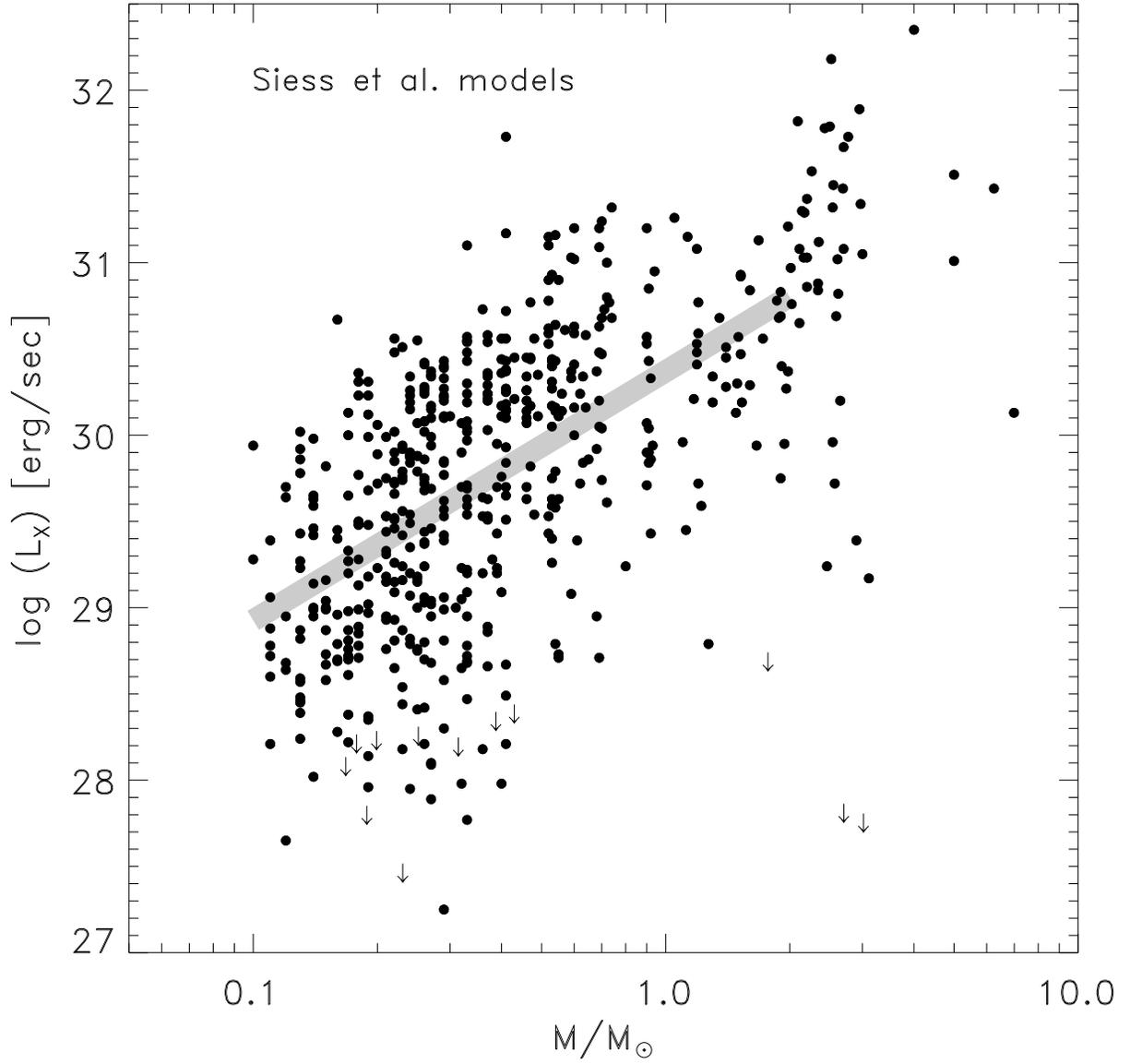}
\figcaption{X-ray luminosity versus stellar mass for the stars
in the optical sample based on masses determined with the PMS models
of \cite{Siess00}.
The thick grey line shows
the linear regression fit to the low-mass ($M \leq 2\,M_\odot$) stars
with the EM algorithm computed with ASURV.
\label{lx_mass_sdf_ps.fig}}
\end{figure}

Next we considered the relation between X-ray luminosity and stellar mass
(Fig.~\ref{lx_mass_sdf_ps.fig}).
As described in \cite{Getman05a}, the stellar masses
listed in the COUP tables were derived from the \citet[][SDF hereafter]{Siess00} PMS models. 
It is well known
that mass estimates from PMS evolutionary models
are subject to significant uncertainties;
different PMS models and/or temperature scales
can lead to differences by as much as a factor of $\sim 2$
in the mass estimates (for  detailed investigations of these
uncertainties see, e.g., \citealp{Luhman99} or \citealp{Hillenbrand04}).
As a test to what extent the $L_{\rm X}\leftrightarrow M$ relation
is dependent on the choice
of the PMS model\footnote{Our comparison here is restricted to
the PMS models of \citet{Siess00} and
\citet{Palla99} because these two models cover particularly wide
ranges of stellar masses; this does not imply that we consider these specific
models to be ``better'' than other models.},
we compared
the relation found for the masses derived from the SDF models
to those based on stellar masses estimated from the PMS models
of \citet[][PS hereafter]{Palla99}.  Note that
the masses determined from these two sets of models agree very well
with each other for objects with $M > 0.4\,M_\odot$, but below
 $0.4\,M_\odot$ the PS models yield
systematically lower masses than the SDF models.

For both sets of stellar masses a clear correlation is found
between X-ray luminosity and mass.
For the low-mass ($M \leq 2\,M_\odot$) stars,
the SDF models lead to an EM linear regression fit of
$\log\left(L_{\rm X}\,[{\rm erg/sec}]\right) = 30.37(\pm0.06) + 1.44(\pm0.10)\times\log\left(M/M_\odot\right)$
 with a standard deviation of 0.65, whereas the
PS models yield a somewhat shallower relation of
$\log\left(L_{\rm X}\,[{\rm erg/sec}]\right) = 30.34(\pm0.05) + 1.13(\pm0.08)\times\log\left(M/M_\odot\right)$ with a standard deviation of 0.64.
From this exercise we conclude that the detailed shape of the
$L_{\rm X}\leftrightarrow M$ correlation does depend on the PMS model used,
but the general dependence is independent of the choice of the model.

The power-law slopes we find here for the ONC TTS 
are considerably lower than those found for the TTS
in the Chamaeleon star forming region \citep[slope $=3.6 \pm 0.6$ in the
mass range $0.6-2\,M_\odot$;][]{Feigelson93}
and the very young stellar cluster IC~348 \citep[slope $=2.0 \pm 0.2$
in the mass range $0.1-2\,M_\odot$;][]{PZ02}
or than that derived for 
M-type field stars \citep[slope $= 2.5 \pm 0.5$ in the mass range $0.15-0.6\,M_\odot$;][]{Fleming88}.  
The differences in the slopes are in part due to differences in the 
considered mass ranges and in the methods to estimate stellar masses.
Another factor may be that many of the previous studies had to deal with
large numbers of X-ray upper limits for undetected
very-low mass stars, which perhaps caused the typical X-ray luminosities
of these very-low mass stars to be underestimated.

\bigskip

\begin{figure}[p]
\plotone{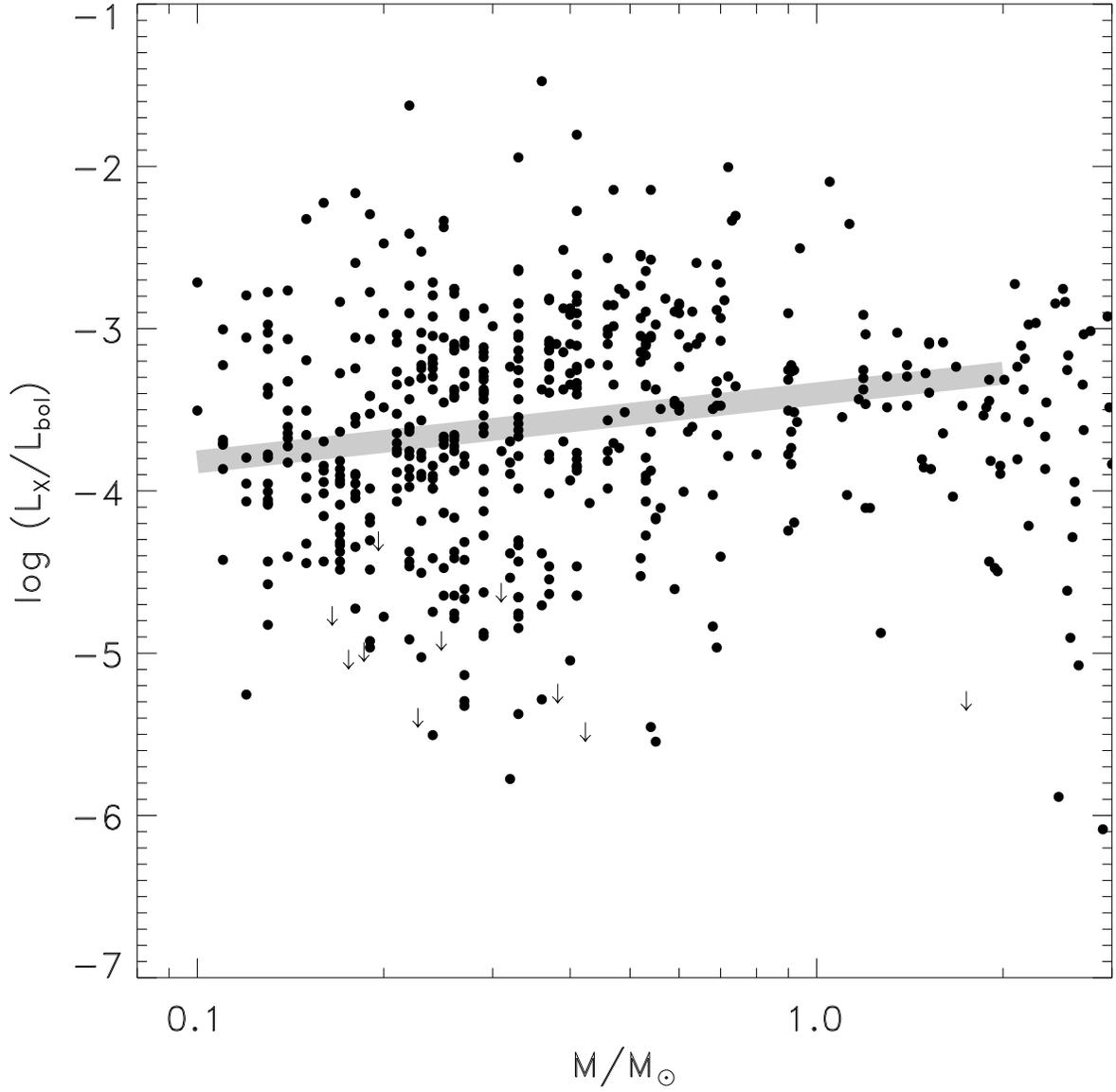}
\figcaption{Fractional X-ray luminosity versus stellar mass for the
low-mass  stars in the COUP optical sample. 
The line shows
the linear regression fit for the low-mass stars ($M \leq 2\,M_\odot$)
with the EM algorithm computed with ASURV.
\label{lxlb_mass.fig}}
\end{figure}

Next we consider the fractional X-ray luminosity as a function of mass.
In Fig.~\ref{lxlb_mass.fig} we show the 
$L_{\rm X}/L_{\rm bol} \leftrightarrow M$ relation for low-mass stars.
The statistical tests in ASURV reveal a very shallow, but nevertheless
highly significant ($P(0) < 10^{-4}$) correlation between
fractional X-ray luminosity and mass for low-mass ($M < 2\,M_\odot$) objects.
The linear regression fit with the EM algorithm  yields
$\log\left(L_{\rm X}/L_{\rm bol}\right) = -3.40(\pm0.06) + 0.42(\pm0.11)\times\log\left(M/M_\odot\right)$ 
with a standard deviation of 0.69~dex.

We also used this relation to illustrate the influence of X-ray
variability during the COUP observation on the resulting correlations.
If we consider the {\em characteristic} X-ray luminosities 
derived from the MLB lightcurve analysis  rather than the {\em average} X-ray 
luminosities, we find a very similar relation with 
$\log\left(L_{\rm X, char}/L_{\rm bol}\right) = -3.56(\pm0.05) + 0.40(\pm0.10)\times\log\left(M/M_\odot\right)$ 
with a standard deviation of 0.64~dex.
This test shows that the use of the characteristic rather than the 
average X-ray luminosities
decreases the values of the fractional X-ray luminosities slightly,
but the power-law slopes for the $L_{\rm X}/L_{\rm bol}\leftrightarrow M$
correlations are nearly identical, and the scatter in the
correlation diagrams is only very slightly smaller.
From this exercise, we conclude that the correlations do not
significantly depend on the choice of the average or characteristic X-ray
luminosities.

\subsection{X-ray activity and stellar effective temperature}

\begin{figure}[p]
\plotone{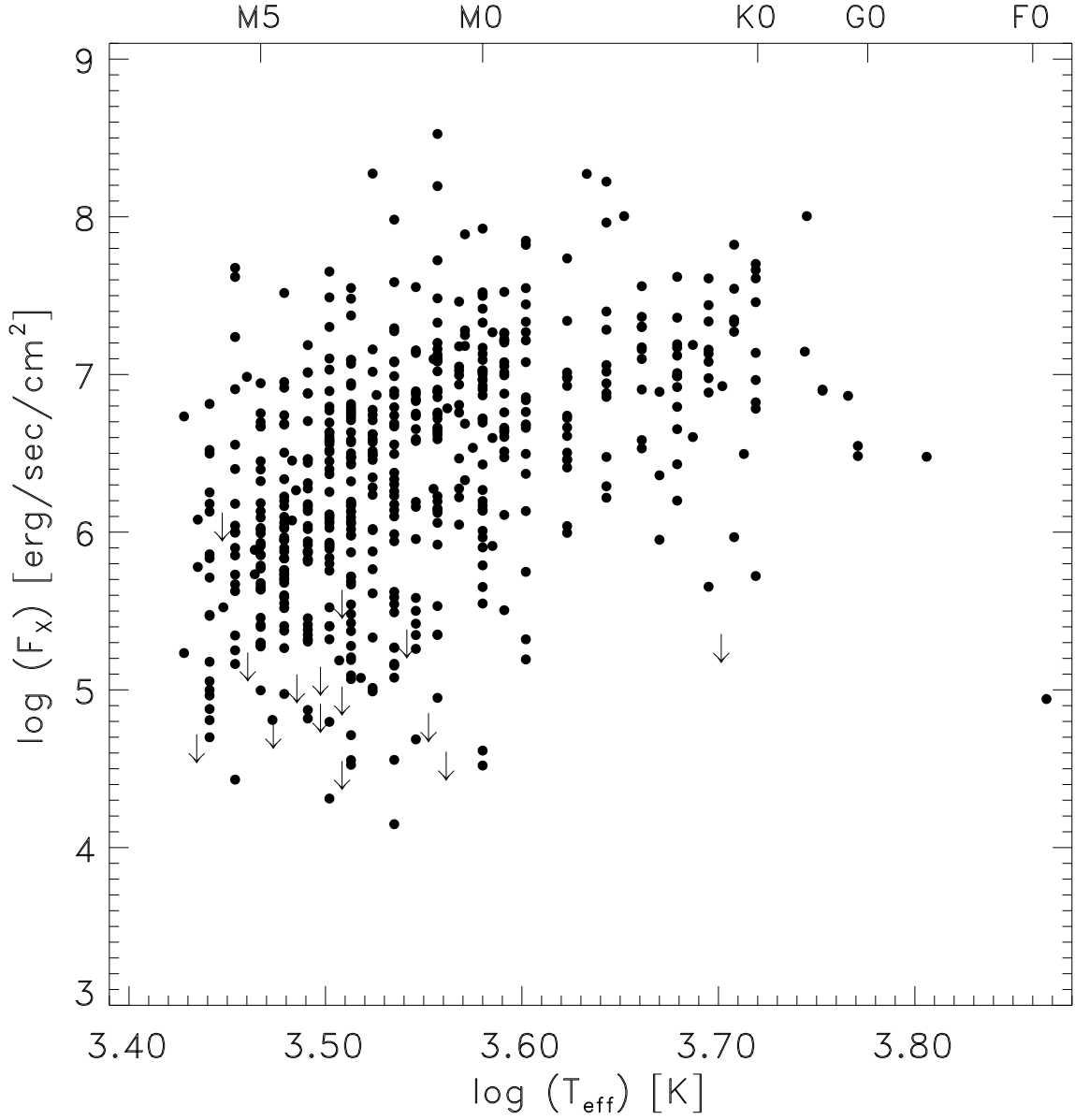}
\figcaption{X-ray surface flux 
versus effective temperature for
the TTS in the optical sample.
\label{fx_teff.fig}}
\end{figure}

In Fig.~\ref{fx_teff.fig} we show the 
X-ray surface fluxes (i.e.~X-ray luminosities divided by
the stellar surface area) of the ONC TTS  plotted
versus their effective temperatures.
Nearly all stars have X-ray surface fluxes in the range
$10^4-10^8\,{\rm erg/cm^2/sec}$,
which corresponds nicely to the minimum and maximum X-ray surface flux 
found for different structures in the solar corona
(where coronal holes and the background corona show X-ray fluxes around
$10^4\,{\rm erg/cm^2/sec}$, while active regions show fluxes
up to $10^8\,{\rm erg/cm^2/sec}$); the similarity of the X-ray surface flux 
ranges found for late-type stars and for different constituents of the
solar corona has already been noted, e.g.,~in \cite{Schmitt97} or
\cite{Peres04}.
The plot also shows a strong decline of the X-ray surface flux
with effective temperature among the M-type stars.
This dependence is much more pronounced than the very shallow relation
between $L_{\rm X}/L_{\rm bol}$ and the stellar mass discussed above.
This effect can be understood if one recalls that
the surface flux and the fractional X-ray luminosity are related
to each other by 
$F_{\rm X} \propto T_{\rm eff}^4 \times \left(L_{\rm X}/L_{\rm 
bol}\right)$. Therefore, $F_{\rm X}$ decreases with decreasing
$T_{\rm eff}$ for constant $L_{\rm X}/L_{\rm bol}$.

\subsection{Comparison to main-sequence stars}

For a meaningful comparison of the ONC TTS to main-sequence (MS) stars,
it is important to keep in mind that our
ONC TTS sample is an {\em optically selected and representative sample}
of cluster members. For a proper comparison we therefore have to use
an optically selected (not X-ray selected) sample of MS stars.
A well suited comparison sample is the
NEXXUS database \citep{Schmitt04}, which provides updated ROSAT X-ray and
optical data (including accurate HIPPARCOS parallaxes) for nearby field stars.
It contains volume-limited 
samples for G-type ($d_{\rm lim} = 14$~pc), K-type ($d_{\rm lim} = 12$~pc),
and M-type ($d_{\rm lim} = 6$~pc) stars with detection rates of more
than 90\%.
The NEXXUS tables were kindly provided to us by the authors; they
list $M_V$, $B-V$, $L_{\rm X}$ in the $0.1-2.4$~keV ROSAT band, and
the X-ray surface flux $F_{\rm X}$. We used the data for the
43 G-type stars (including 4 non-detections),
the 54 K-type stars (including 2 non-detections), and the 79 M-type
stars (including 5 non-detections).
Bolometric luminosities, effective temperatures, and masses of the
stars were estimated by interpolation using MS relationships
of these quantities with the absolute magnitude $M_V$.

When comparing the NEXXUS data to our COUP data
 one has to take into account the different energy
bands for which X-ray luminosities were computed.
The NEXXUS X-ray luminosities are given for the $0.1\!-\!2.4$~keV
ROSAT band, and for comparison with our COUP results we have to transform
these luminosities into $0.5\!-\!8$~keV band.
The transformation factor can be calculated with PIMMS and
depends on the X-ray spectrum; 
since we can assume thermal plasma spectra, the transformation
mainly depends on the plasma temperature.
For the NEXXUS stars, the count-rate to luminosity transformation factor
used by \cite{Schmitt04} assumes a plasma temperature of $\sim 2.5$~MK,
and the corresponding energy band correction factor is 0.33~dex.
In the following comparisons we also consider the X-ray properties 
of our Sun.  For this, we use here the ROSAT-band X-ray luminosity range of
 $\log\left(L_{\rm X}\,[{\rm erg/sec}]\right) = 26.8-27.9$
based on the results of \cite{Judge03} for the
activity range of a typical solar cycle. Assuming a plasma temperature
of 2~MK, the flux in the $0.5-8$~keV band is 0.48~dex lower than
that in the $0.1-2.4$~keV band.

\begin{figure}[p]
\plotone{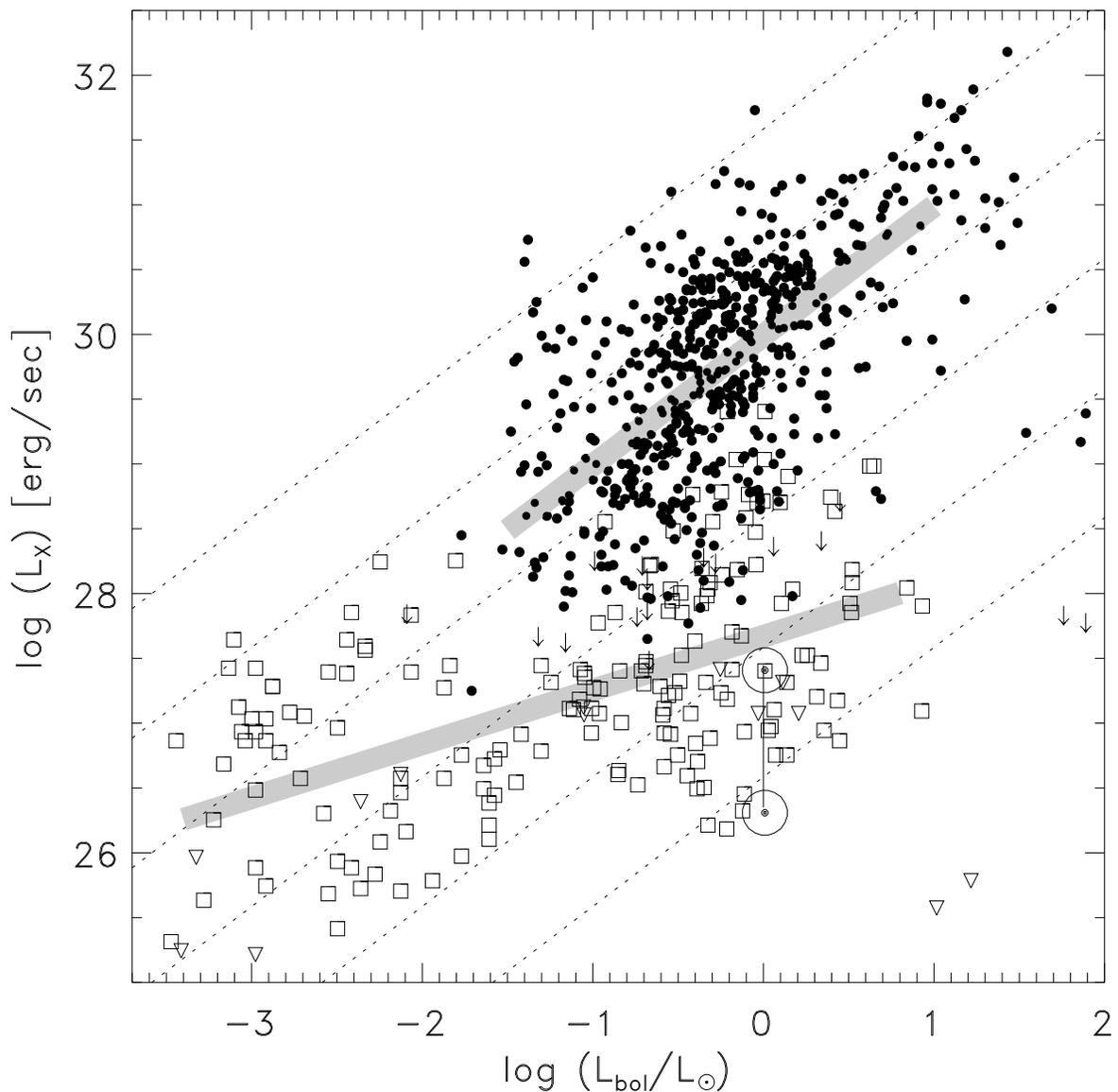}
\figcaption{X-ray luminosity versus bolometric luminosity for the stars
in the COUP optical sample (solid dots, arrows for upper limits)
and for the NEXXUS sample of nearby field stars (open squares, triangles for
upper limits).
The dotted lines mark $\log\left(L_{\rm X}/L_{\rm bol}\right)$ ratios of $-2$,
$-3$, $-4$, $-5$, $-6$, and $-7$.
The thick grey lines
show the linear regression fits with the EM algorithm computed with ASURV
for these two samples.
\label{lx_lb_nex.fig}}
\end{figure}

Figure \ref{lx_lb_nex.fig} compares the $L_{\rm X}\!\leftrightarrow\!L_{\rm bol}$
relations for the COUP optical sample to that for the NEXXUS sample
of field stars.
A clear correlation between
X-ray and bolometric luminosity is not only seen for the ONC TTS, but also
for the NEXXUS field stars. The tests in ASURV
show that the correlation for the NEXXUS stars is significant
$(P(0) < 10^{-4})$; the linear regression fit with the EM algorithm
for the G-, K-, and M-type stars in the NEXXUS sample 
yields a power-law slope of $0.42\pm0.05$, which is much shallower
than the slope found for the $L_{\rm X}\!\leftrightarrow\!L_{\rm bol}$
correlation for the COUP stars ($1.04\pm0.06$).

\begin{figure}[p]
\plottwo{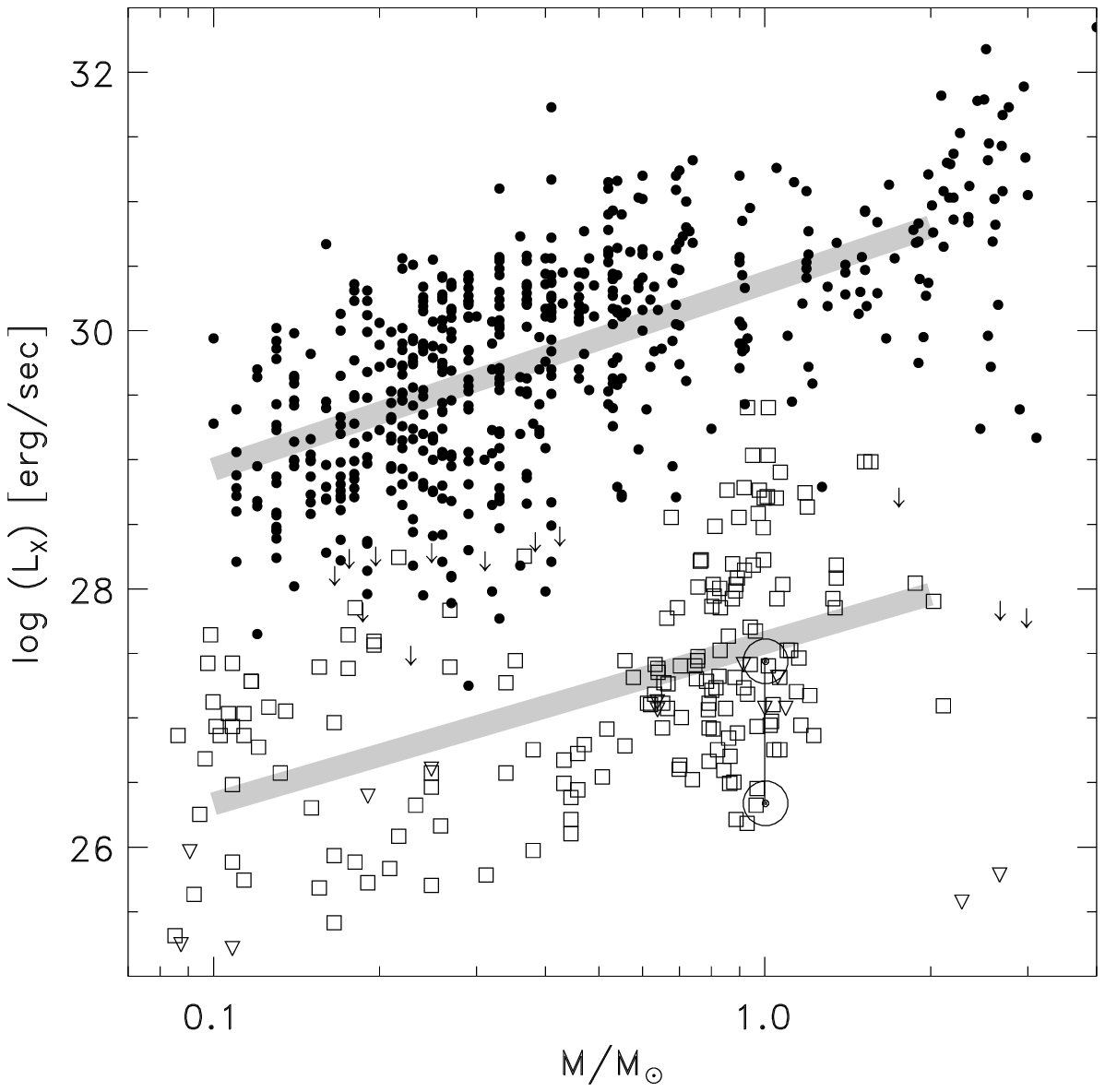}{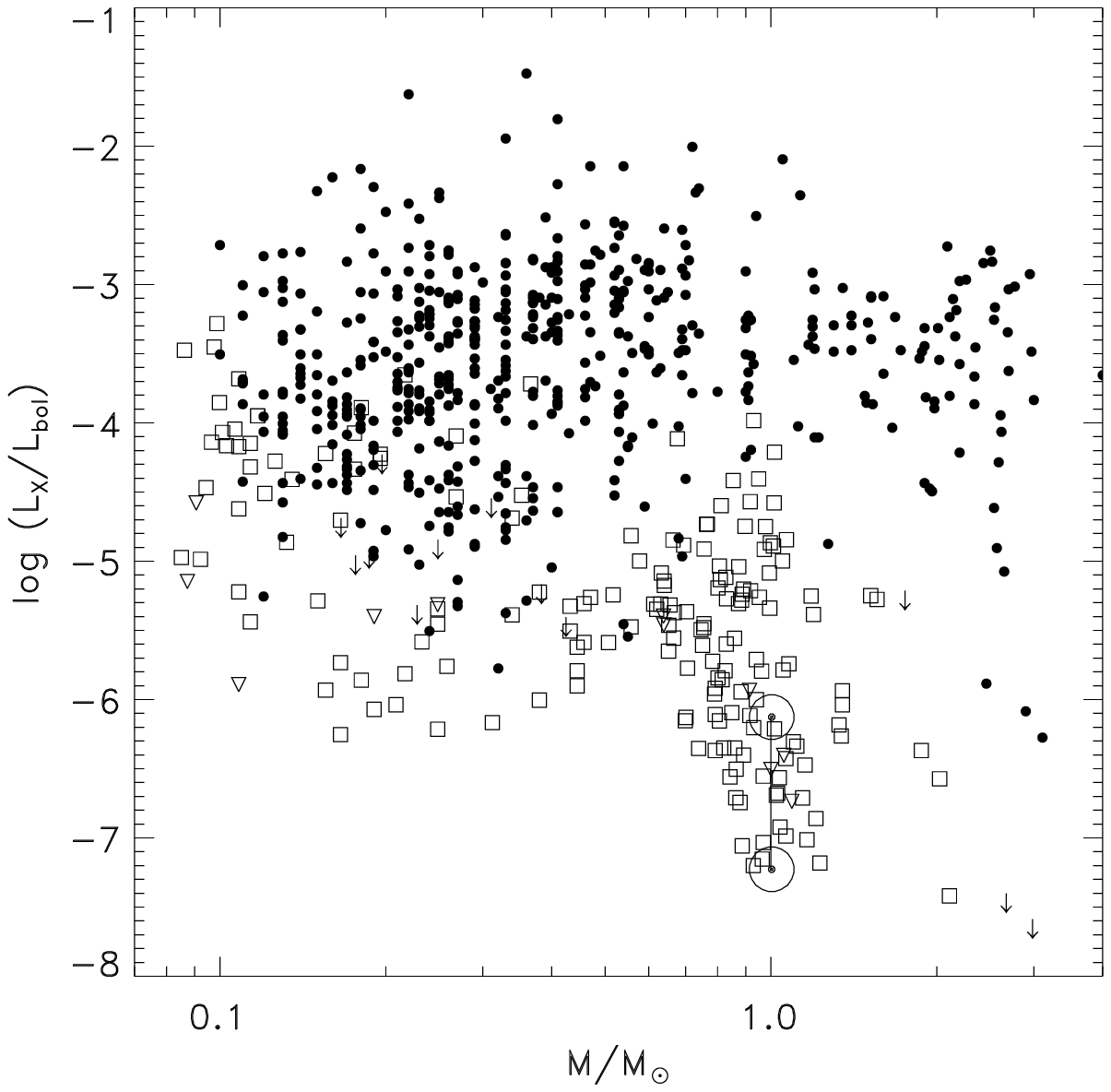}
\figcaption{Absolute (left) and fractional (right)
X-ray luminosity versus stellar mass for the stars
in the COUP optical sample (solid dots, arrows for upper limits)
based on masses determined with the PMS models
of \cite{Siess00}, and for the NEXXUS sample of nearby field stars
(open squares, triangles for upper limits).
The thick grey lines in the $L_{\rm X} \leftrightarrow M$ correlation
show the linear regression fits
with the EM algorithm in ASURV
for the low-mass ($M \leq 2\,M_\odot$) stars in these two samples.
\label{lx_mass_ms.fig}}
\end{figure}

Figure \ref{lx_mass_ms.fig} compares the $L_{\rm X}\!\leftrightarrow\!M$
relations for the COUP optical sample to that for the NEXXUS sample
of field stars.
It is interesting to see that there is a clear correlation between
X-ray luminosity and mass for the NEXXUS field stars. The tests in ASURV
show the X-ray luminosity and mass are clearly correlated
$(P(0) < 10^{-4})$; the linear regression fit with the EM algorithm
for objects in the mass range $0.08 - 2\,M_\odot$
yields
$\log\left(L_{\rm X}\,[{\rm erg/sec}]\right) = 27.58(\pm0.07) + 1.25(\pm0.15)\times\log\left(M/M_\odot\right)$
with a standard deviation of 0.77.
The corresponding correlation for the TTS in the COUP optical sample 
yielded a power-law slope of $1.44\pm0.10$, which is consistent to the
slope for the field stars within the uncertainties.
The similarity of the slopes found in the $L_{\rm X}\!\leftrightarrow\!M$
relations for the ONC TTS and the field stars
may indicate that the relation between stellar mass and X-ray 
luminosity is more fundamental than that between
bolometric and X-ray luminosity.

The plot of fractional X-ray luminosities against stellar masses
(Fig.~\ref{lx_mass_ms.fig}) shows that some of the very-low mass field stars
reach similar activity levels as the TTS.
The solar-mass field stars, on the other hand, are typically much less 
X-ray active than their young COUP counterparts.
The different activity levels of the field stars as a function of mass
can be understood as a consequence
of the activity-rotation relation for MS stars: 
many of the very-low mass field stars are rapid rotators, thus show
high levels of X-ray activity,
while most solar-mass field stars rotate quite slowly, therefore
displaying lower activity levels.
For the COUP stars, on the other hand, we show in 
Section \ref{rotation.sec} that all stars with known rotation period
rotate more rapidly than the Sun, and that their X-ray activity is
probably unrelated to their rotation period.

\section{X-ray emission, rotation, and convection\label{rotation.sec}}

\subsection{The activity--rotation relation for the TTS\label{lxlb_prot.sec}}

For MS stars,
the well established correlation between
fractional X-ray luminosity and rotation period
\citep[e.g.][]{Pallavicini81, Pizzolato03} constitutes the main argument for
solar-like dynamo mechanism as the origin of their X-ray activity.
As already noted in the introduction, the existence of a similar
activity\,--\,rotation  relation could not be unambiguously proven for
PMS stars, mainly due to a lack of statistical power in the
underlying data
(in most studies the sample sizes were
too small for statistically significant conclusions to be drawn).
The previous $Chandra$ ONC studies \citep{Feigelson02a, Flaccomio03b},
however, provided strong evidence that the TTS do {\em not} follow
the activity\,--\,rotation relation for MS stars.

\begin{figure}[p]
\plotone{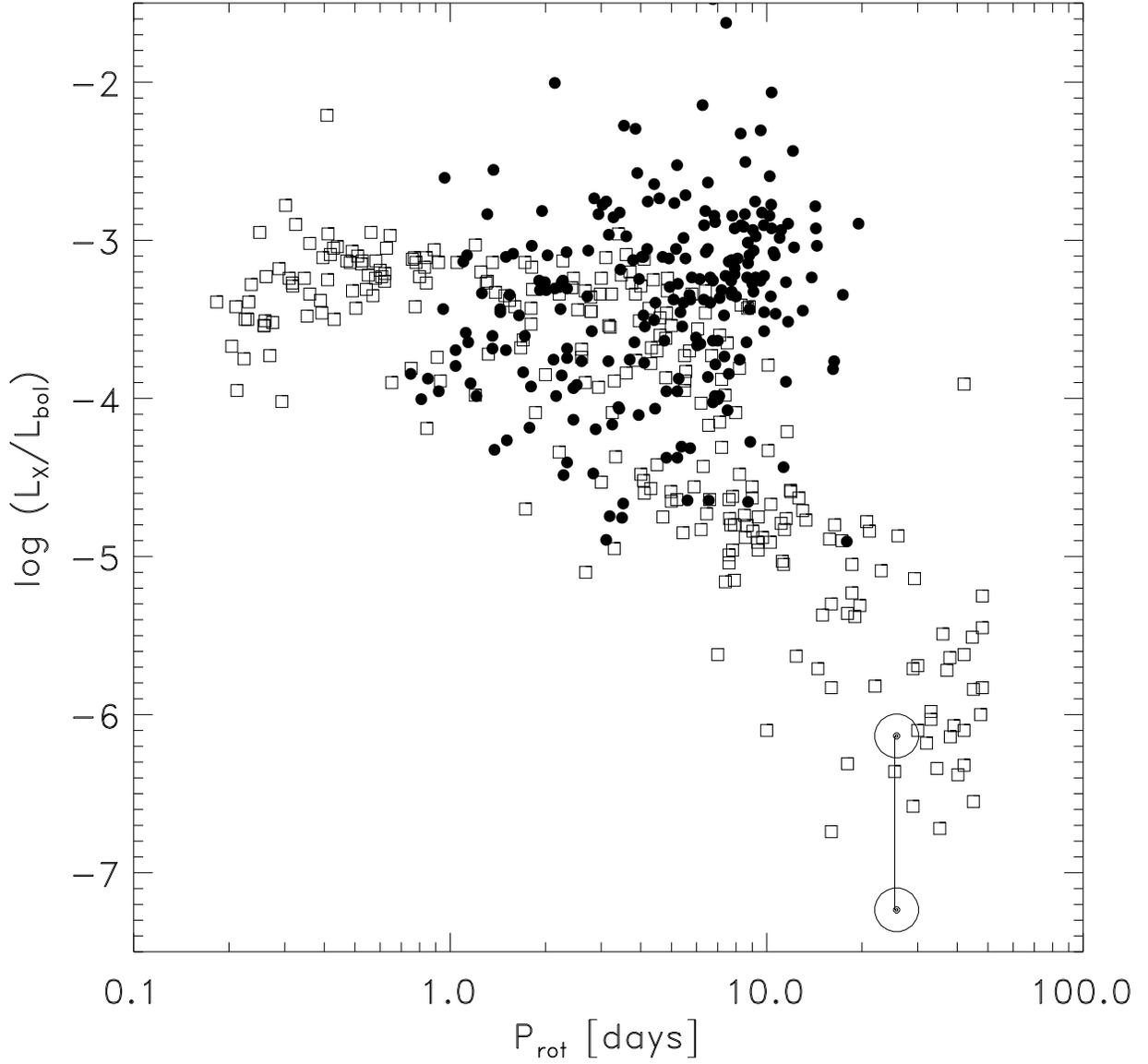}
\figcaption{Fractional X-ray luminosity versus rotation period.
This plot compares the ONC TTS (solid dots)
to data for MS stars from \cite{Pizzolato03} and \cite{Messina03}
(open boxes) and the Sun.
\label{lxlb_prot.fig}}
\end{figure}

 Table 9 in \citet{Getman05a} lists rotation periods for
158 stars in our COUP optical sample.
Considering also the additional rotation data as listed
in \cite{Flaccomio05}, rotation periods from 
photometric monitoring are available for
228 stars in our optical TTS sample (169 M-type stars, 58 K-type stars,
and one G-type star). 
In Fig.~\ref{lxlb_prot.fig} we plot the fractional X-ray luminosity versus 
rotation period for these stars, and compare them to data for MS
stars. It is rather obvious that the COUP stars do not follow the 
well established activity-rotation relation shown by the MS stars,
i.e.~increasing activity for decreasing
rotation periods followed by saturation at $L_{\rm X}/L_{\rm bol} \sim 10^{-3}$
for the fastest rotators. 
A statistical analysis reveals for the  COUP stars a
correlation between $L_{\rm X}/L_{\rm bol}$ and $P_{\rm rot}$ rather
than the  anti-correlation seen for the MS stars:
the linear regression analysis with SLOPES \citep{Isobe90, FB92} yields a
bisector regression fit
of the form $\log\left(L_{\rm X}/L_{\rm bol}\right) = -4.21(\pm 0.07) +
1.27(\pm0.09)\times\log\left(P_{\rm rot}\,[{\rm days}]\right)$.
This correlation is statistically significant;
a Kendall's $\tau$ and Spearman's $\rho$ test give
probabilities of $P(0) = 0.0002$ 
for the null hypothesis that a correlation is not present.
The observed  correlation between $L_{\rm X}/L_{\rm bol}$ and $P_{\rm rot}$ 
is clearly very different from the anti-correlation shown by the MS
stars (where the bisector regression fit yields a slope of $-2.35\pm0.16$
for the sample shown in Fig.~\ref{lxlb_prot.fig} in the period range
1--10 days).
These results do not change significantly if we use the
characteristic rather than the average X-ray luminosities.

Before we consider possible explanations for these findings,
it is important to note that rotation periods are known for
only $\sim 38\%$ of the X-ray detected stars in our optical sample, 
and that this subsample may be biased with respect to its X-ray 
properties. \cite{Stassun04a} studied archival ACIS data of the ONC 
and pointed out that the stars with known rotation period in their sample
show systematically higher X-ray activity than the stars with unknown
periods.
A similar difference is present in our COUP optical sample:
the median fractional X-ray luminosity for the TTS with known
rotation periods 
is at 
$\log\left(L_{\rm X}/L_{\rm bol}\right)= -3.31$, while the median value
 for TTS with unknown periods is $-3.71$. A KS test
gives a probability of $P(0) \ll 10^{-4}$ % only $2.35\times 10^{-8}$ 
for the hypothesis that the
distributions of fractional X-ray luminosities in both samples
are identical, i.e.~the apparent difference of about a factor of
$\sim 2.5$ in X-ray activity is statistically significant. 
This difference is not due to systematic differences in the
basic stellar parameters\footnote{We note 
that the stars with
known periods have systematically higher masses 
($\Delta \log\left(M\right) \sim 0.06$~dex, $P_{\rm KS}(0) = 0.006$)
than the stars without periods.
According to the correlation between  stellar mass and fractional
X-ray luminosity established in Sect.~\ref{relations.sec},
this difference in stellar masses would, however, predict a difference in the
activity level of only
$\Delta \log \left(L_{\rm X}/L_{\rm bol}\right) \sim 0.02$~dex,
much smaller than the observed $0.4$~dex difference between stars with and
without known periods. This effect can therefore not explain 
the difference in the activity levels of stars with and without known
periods.} of the two samples.

The explanation of this difference was discussed in detail by \cite{Stassun04a}:
Rotation periods can only be determined for stars showing sufficiently large
spot related photometric variability. The level of photometric variability,
however,
is related to the level of magnetic activity, and therefore the more
active stars (i.e.~those with higher X-ray luminosities) are
easier targets for a determination of photometric rotation periods, 
while the less active stars (i.e.~those with lower X-ray luminosities) 
show too small photometric variability to allow determination of periods.
A quantitative description of these interrelations for the case of
MS stars has been given by \cite{Messina03}.
It is therefore likely that the COUP stars without known rotation
periods rotate on average slower than the stars with known periods.
This introduces a bias our sample,
since most of the missing stars
(i.e.~those without known rotation periods) have
lower fractional X-ray luminosities and longer rotation periods
than the stars in our sample.
The apparent correlation between X-ray activity and rotation period
may therefore (in part) be due to these selection effects.

Nonetheless, it appears very unlikely that the ONC TTS follow the same
rotation-activity relation as found for MS stars. First,
we note that the effects
of the bias due to unknown rotation periods appear much too small
to transform the strong $L_{\rm X}/L_{\rm bol} \leftrightarrow P_{\rm rot}$
anti-correlation of the MS stars into an apparent positive correlation.
Second, the quite
high fractional X-ray luminosities for most TTS in the period
range between $\sim 7$ and $\sim 15$~days, which are
more than one order of magnitude
higher than the typical values for MS stars, also indicate 
differences between the TTS and the MS sample.
In conclusion, we are confident that the ONC TTS do not follow the same
rotation-activity relation as seen in MS stars, but it remains unclear
whether and how the X-ray activity of the TTS is correlated 
to their rotation periods.

The inability to draw meaningful conclusions from our X-ray and
rotation data may 
be due to the fact that the rotation period is perhaps the wrong variable to 
look at.
Theoretical studies of the solar-like $\alpha-\Omega$ dynamos
show that the dynamo number 
is not directly related to the rotation period, but to more complicated 
quantities such as the radial gradient of the angular velocity and the
characteristic scale length of convection  at the base of the
convection zone.
It can be shown that (with some reasonable assumptions) the dynamo number 
is essentially proportional to the inverse square of the Rossby number $Ro$
\citep[e.g.][]{Maggio87}.
The Rossby number is defined as the ratio of the rotation period to the
convective turnover time $\tau_c$, i.e.~$Ro := P_{\rm rot}/\tau_c$.
For MS stars, the theoretical expectations are well confirmed:
it is well established \citep[e.g.][]{Montesinos01}
that the  stellar activity shows a tighter relationship to the
Rossby number than to rotation period. The shape of the relation
is similar to that of the activity--rotation relation:
for large Rossby numbers, activity rises strongly as 
$L_{\rm X}/L_{\rm bol} \propto Ro^{-2}$ 
until saturation at $L_{\rm X}/L_{\rm bol} \sim 10^{-3}$ is reached
around $Ro \sim 0.1$,
which is followed by a regime of ``supersaturation'' 
for very small Rossby numbers, $Ro \lesssim 0.02$.

The convective turnover time scale is a sensitive function of 
the physical properties in the stellar interior and its
determination requires detailed stellar structure models.
Many studies of stellar activity therefore used
semi-empirical interpolations of $\tau_c$ values as a function of,
e.g.,~$B-V$ color. This may be appropriate for MS stars,
but seems to be insufficient for TTS which have a very different and
quickly evolving internal structure.

\subsection{Computation of convective turnover times for the TTS}

For the computation of convective turnover times a
series of stellar evolution models with masses ranging from $0.065\,M_\odot$
 to $4.0\,M_\odot$ was computed with the Yale
Stellar Evolution Code.
The evolution was assumed to start at the stellar birthline, 
where stars initially become visible objects \citep{Palla93}.
All models used the parameters derived for the standard solar model,
where the initial $X$, $Z$, and the mixing-length ratio were varied until a
solar model at the solar age of 4.55~Gyr has the
observed solar values of luminosity, radius, and $Z/X$ \citep[=0.0244;][]{Grevesse96}.
The model that best matched the solar properties\footnote{We note that 
new precision measurements of elemental abundances 
on the solar surface imply a lower metalicity than previously assumed,
and lead to inconsistencies in theoretical solar models with respect
to the depth of the solar convection zone \citep[see][and reference therein]{Bahcall04}.}
was of $(X,Z)_{\rm initial}=(0.7149, 0.0181)$ and the mixing length ratio
1.7432. These values were then used for all stellar models.

A detailed discussion of the physics used in this study for the construction 
of stellar models can be found in \cite{Yi01}. The most important aspects
are as follows:
The solar mixture was assumed as given by \cite{Grevesse93}.
For $\log T >4$ OPAL Rosseland mean opacities \citep{Iglesias96}
were used, for $\log T < 4$ opacities from \cite{Alexander94}.
The OPAL EOS 2001 equation of state \citep{Rogers96} was used and the 
energy generation rates were set according to \cite{Bahcall92}.
Neutrino losses were taken following \cite{Itoh89}, and for the
helium diffusion the values of  \cite{Thoul94} were used.

\begin{figure}[p]
\plotone{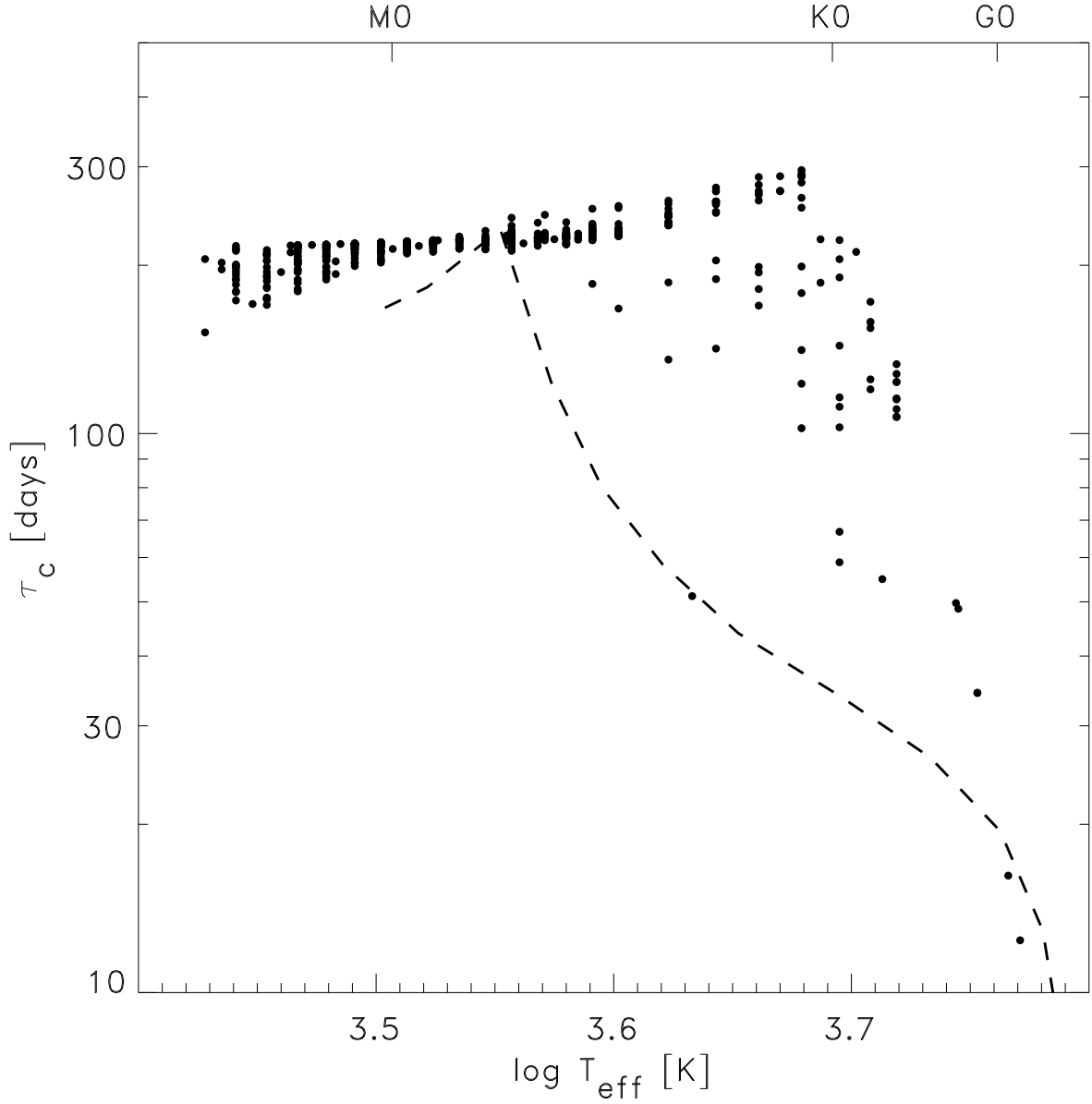}
\figcaption{The local convective turnover time versus spectral type
for stars in the COUP optical sample (solid dots).
The dashed line shows the local convective turnover times
for 4.5~Gyr old stars.
\label{tauc_teff.fig}}
\end{figure}

Since the dynamo action is believed to take place
at the base of the convection zone, anchored in the radiative layers
just below the convective interface, the convective turnover time of the
deepest part of the convection zone is the most relevant in the
evaluation of the Rossby number.
Our knowledge of stellar convection is too limited to calculate
'correct' convection turnover times, because the characteristic length scales,
as well as the velocities, are not well known. Even when one decides to
resort to the mixing length approximation, there are still
uncertainties: the mixing length ratio is assumed to be the same for all
stars with different masses and/or at different evolutionary stages,
which is probably not fully correct.  However, for the convection near the
base of the convection zone where the temperature gradient is for all
practical purposes adiabatic, the mixing-length approximation is known
to provide an adequate description of convection at least in an average
sense \citep{Kim96}.
For the characteristic timescale of convective overturn, the convective
turnover time (i.e.~the local mixing length divided by the local
velocity) was calculated at each time step, which was determined at
a distance of one-half the mixing length above the base of the surface
convection zone\footnote{Note that for fully convective stars 
the  base of the convection zone is the center.} \citep{Gilliland86, Kim96}.
The convective turnover times were determined for the 
stars in the optical sample 
according to the model that represents their corresponding
state in the HR-diagram, i.e.~reproduces the 
$\left(T_{\rm eff}, L_{\rm bol}\right)$ values. The resulting values
are shown in Fig.~\ref{tauc_teff.fig}.
Note that the convective turnover times of the TTS 
are much larger (up to factors of $\sim 8$) than those in MS 
stars, and depend strongly on the effective temperatures and 
ages of the stars.

\subsection{The activity -- Rossby number relation for the TTS}

\begin{figure}[p]
\plotone{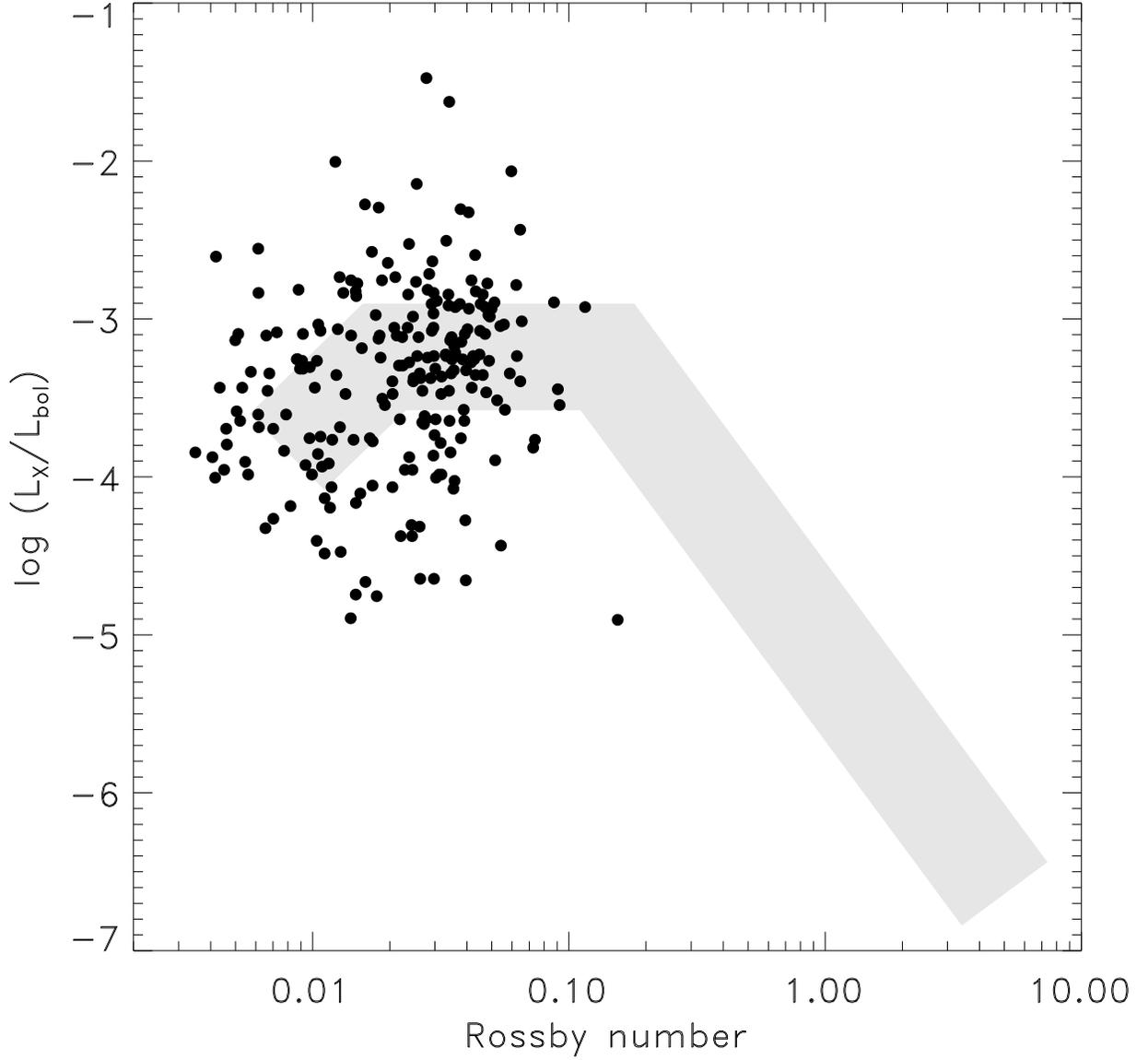}
\figcaption{Fractional X-ray luminosity versus Rossby number
for the COUP stars.
The grey shaded area shows the relation and the width of its
 typical scatter found for MS stars \citep[from][]{Pizzolato03}.
\label{lxlb_rossby.fig}}
\end{figure}

The Rossby numbers for the ONC TTS were computed by dividing their
rotation periods
by the values for their local convective turnover time as derived above.
The plot of fractional X-ray luminosity versus Rossby number in
Fig.~\ref{lxlb_rossby.fig} shows no strong relation between these
two quantities.
All TTS have Rossby numbers $< 0.2$ and therefore 
are in the saturated or super-saturated
regime of the activity -- Rossby number relation for MS stars.
To search for indications of super-saturation,
we compared the fractional X-ray luminosities
of the TTS in the saturated ($0.1 > Ro > 0.02$) and super-saturated
($Ro \leq 0.02$) regimes. Indeed, we found a slightly lower median
$\log\left(L_{\rm X}/L_{\rm bol}\right)$ of $-3.63$ for the TTS 
in the super-saturated regime as compared to $-3.43$ for 
those in the saturated regime;
a KS test gives a probability of 
$P(0) = 0.031$ for the assumption
of equal $L_{\rm X}/L_{\rm bol}$ distributions  in both samples,
i.e.~the difference is significant at the 97\% level.
Thus, the fractional X-ray luminosities of the TTS show a qualitatively
similar relation to their Rossby numbers as is found for MS stars.

A remarkable difference between our TTS sample and the data for
MS stars is the very wide dispersion of fractional X-ray luminosities
at a given Rossby number in our TTS data. The scatter extends over
about three orders of magnitude, and even if we
consider the characteristic (i.e.~flare-cleaned) rather than the
average X-ray luminosities, the scatter is only very slightly smaller.
This large scatter is in strong contrast to
the tight relation found
for MS stars, where the scatter in $\log\left(L_{\rm X}/L_{\rm bol}\right)$
at a given Rossby number is only about $\pm0.5$~dex \citep[e.g.,][]{Pizzolato03}.

To conclude, we find that the X-ray activity of the
ONC TTS may depend on Rossby numbers in a similar way to what is found 
for MS stars,
but the large scatter of X-ray activity at any given Rossby number suggests
that additional factors are important for the level of  X-ray activity.

\section{Possible explanations for the wide scatter in the correlations\label{variability.sec}}

All the correlations between the X-ray activity and other stellar parameters
show a very large scatter, often exceeding three orders of magnitudes.
Three obvious reasons for the presence of scatter are 
uncertainties in the variables, X-ray variability, and 
unresolved binaries.
Can these effects account for the wide scatter seen in the correlations?
We first consider X-ray variability.

Most of the COUP sources show strong variability in their
lightcurves. It is well known
that variability is a common feature in the X-ray emission 
from TTS, active stars, and also our Sun. 
The degree of variability is a function of the timescales.
For active MS
stars and the Sun,
short term variability (timescales of minutes to hours)
is usually dominated by flares which can cause large
variations  (sometimes exceeding factors of 10).
On timescales from days to weeks, the typical variations
are about a factor of 3 and up to 10, while
the typical variability on timescales from months to years is 
a factor of about $3-4$ \citep[e.g.][]{Micela2003, Orlando04}.
Furthermore, indications for stellar X-ray activity cycles 
have been found for some stars; these cycles can induce more than
one order of magnitude variability over a few years \citep[e.g.,][]{Favata04}.

In previous, generally much shorter X-ray observations of
TTS, the large scatter was often assumed as being probably due in part 
to X-ray flaring. The COUP data provide
two important pieces of information in this respect.
First,  the influence of individual flares 
on the average X-ray luminosity is strongly smoothed out by the long
time basis of the COUP observation;
the scatter seen in the correlations based on our COUP data 
should therefore be much smaller than what would be found from
shorter, snapshot-like observations. Second, the use of the
characteristic X-ray luminosities, which effectively exclude periods of
flaring from the lightcurves,  should 
further reduce the scatter, if short-term X-ray variability were the main reason
for the large scatter in the correlation diagrams.
These expectations are, however, not confirmed in our data:
We find that the scatter in the COUP (i.e., 10 day average) 
X-ray luminosities of the ONC TTS, e.g.~as
a function of bolometric luminosity or mass, is very similar to that
found in the correlations based on the previous 23~hr ACIS observation
\citep{Feigelson02a}. Also, the use of characteristic rather than
average X-ray luminosities reduces the scatter only marginally.
This suggests that variability on time-scales between
$\sim 1$ and $\sim 10$~days
is {\em not} the main source of the
large scatter in the correlations between X-ray activity and 
other stellar parameters.

What about variability on longer timescales?
 We can investigate
the variations on timescales of several years 
by comparison of the COUP data (obtained in January 2003)
to the previous 23 hour ACIS
observations of the ONC (obtained in October~1999/April~2000) 
given by \cite{Feigelson02a}.
For 515 of the COUP detected stars in our optical sample X-ray luminosities
based on the 23 hour observations are listed by \cite{Feigelson02a}.
 Since the spectral fitting 
procedure used in the analysis of the 23 hr data is not identical to 
that used for the COUP data, we compare here the {\em observed} X-ray
luminosities, $L_{\rm t}$, without extinction correction, which just
give the integral of the observed flux over the spectrum.
We find
 a  good agreement of the luminosities from the two observations
separated by more than 3 years: 
the median absolute deviation from equal luminosities 
is only 0.31 dex, corresponding to just a factor
of $\sim 2$. 
This demonstrates that the X-ray luminosities of most ONC TTS 
vary only slightly on timescales of several years. The observed level of
variability cannot account for the large scatter seen in the correlation 
diagrams.

Now we try to quantify the uncertainties of the variables in the 
correlation diagrams.
According to H97, the uncertainties in $\log\left(L_{\rm bol}\right)$ are
$\sim 0.2$~dex. The uncertainties in the stellar mass estimates
are probably of a similar magnitude.
The uncertainties of the X-ray luminosities, derived from the spectral fits,
are difficult to determine because the spectral models are highly non-linear
and the individual spectral parameters are often
strongly correlated. Furthermore, ambiguities can occur when
two qualitatively different spectral models give
similarly good fits. We therefore assume the typical 
random uncertainties of $\log\left(L_{\rm X}\right)$ to be similar
to the uncertainties in the emission measures derived in the spectral fits,
i.e.~about $0.15$~dex. Note, however, that some sources may be affected by 
systematic errors, which may well exceed this level.

Finally, we consider the effect of unresolved binaries.
The presence of an unresolved companion causes an overestimation of both,
the bolometric and the X-ray luminosity.
However, if X-ray and bolometric luminosity are related linearly 
(as our data suggest), unresolved companions should only shift the
position of a star in the $L_{\rm X}\leftrightarrow L_{\rm bol}$ diagram
along the correlation line and not increase the scatter.
The stellar mass
determined by comparison with PMS tracks depends (in the case of low-mass 
stars) mainly on the measured spectral type; as the combined optical spectrum
of a binary system is dominated by the light of the primary component,
the inferred mass is that of the primary, whereas the observed X-ray
luminosity is the sum for primary and companion. 
If X-ray luminosity and stellar mass are correlated,
the overestimation of the X-ray luminosity should be at most a factor of
two\footnote{There may, however, be larger effects in special cases. 
For example, in the case of a binary system in which both components show
very different amounts of extinction, a fit to the combined
X-ray spectrum may easily lead to wrong parameters.};
the typical shift in $\log\left(L_{\rm X}\right)$
 depends on the distribution of mass ratios in the binary systems 
(which is not well known), but is presumably about $0.2-0.3$~dex among
the low-mass stars.

The combined effect of the variability, the uncertainties in the variables,
and unresolved binaries
should therefore produce a scatter of roughly $\pm(0.4-0.5)$~dex.
This is considerably less (by about a factor of two) 
than the scatter observed in the correlation 
diagrams, where the standard deviations are typically 
$\sim 0.7$~dex. 
A large fraction
of the scatter in the correlations must therefore be due to other reasons,
most likely due to intrinsic differences 
in the X-ray activity levels of the TTS. In \S\ref{accretion.sec} 
we will show that the large scatter is probably related  to the
influence of accretion on the X-ray properties.

\section{X-ray plasma temperatures\label{spectra.sec}}

As described in G05, the X-ray spectra of the COUP sources were fitted
with single-temperature  or (in most cases) two-temperature thermal
plasma models plus absorption.
We are fully aware that these relatively simple models are not ``perfect",
since it is well known that the coronae of active stars are generally
not monothermal and can usually not be considered as to consist of 
just two different temperature components 
\citep[e.g.][]{Brickhouse2000, Sanz03}. 
Nevertheless, this approach is justified because the
purpose of our analysis was to characterize the coronal temperatures of the 
COUP stars in a homogenous way using a parsimonious model, rather
than to perform a detailed investigation
of the temperature structure of those sources (which will be the topic
of separate studies).
As demonstrated for example by 
Peres et al.~(2000), fits 
of simulated spectra based on continuous temperature distributions
with simple one- or two-temperature models
usually yield temperatures near the peaks of the
underlying temperature distribution.
We therefore assume that the temperatures derived from the fits
reflect some kind of a
``characteristic'' temperature, which then can be related to other 
stellar parameters.

\begin{figure}[p]
\plottwo{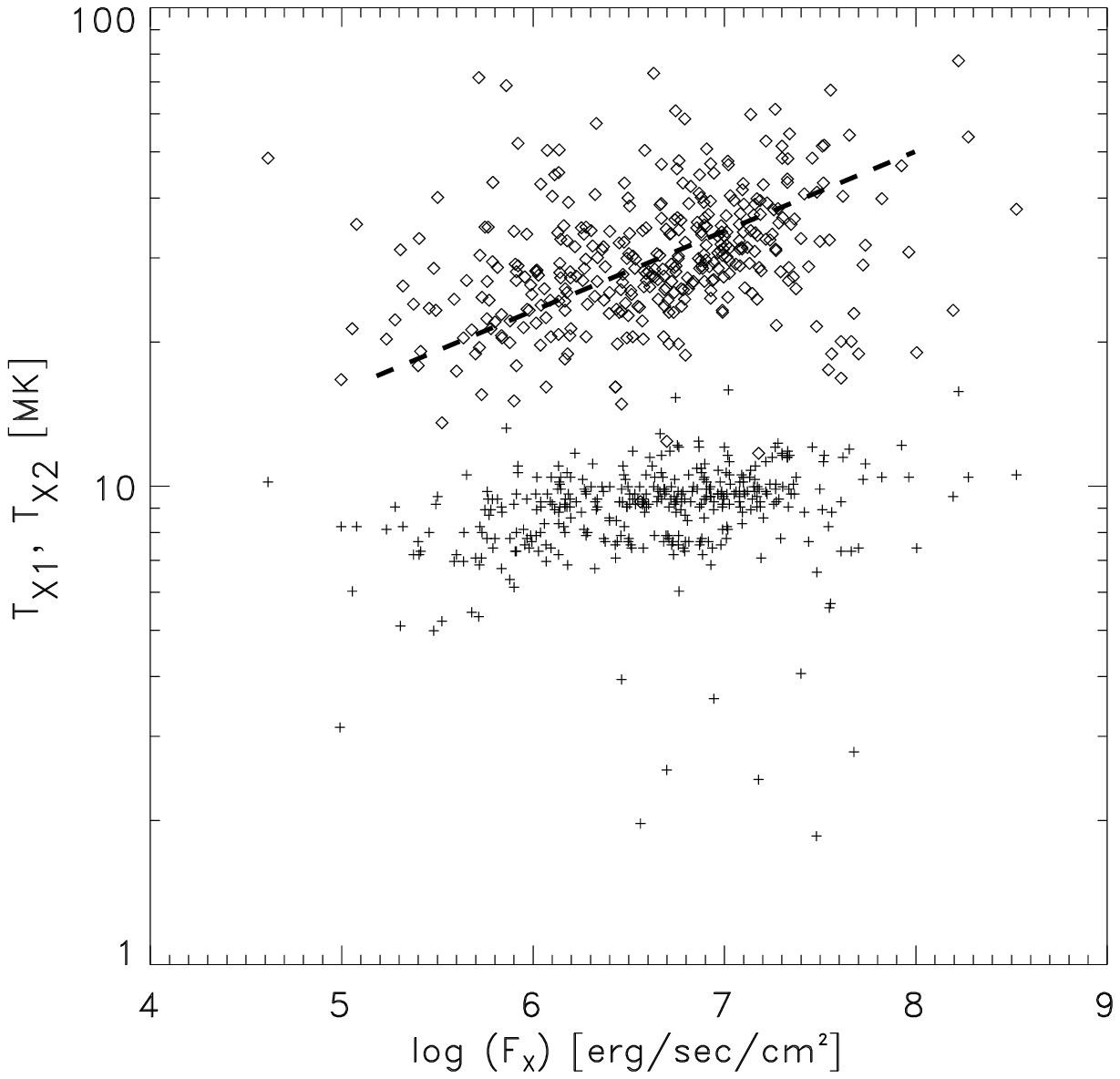}{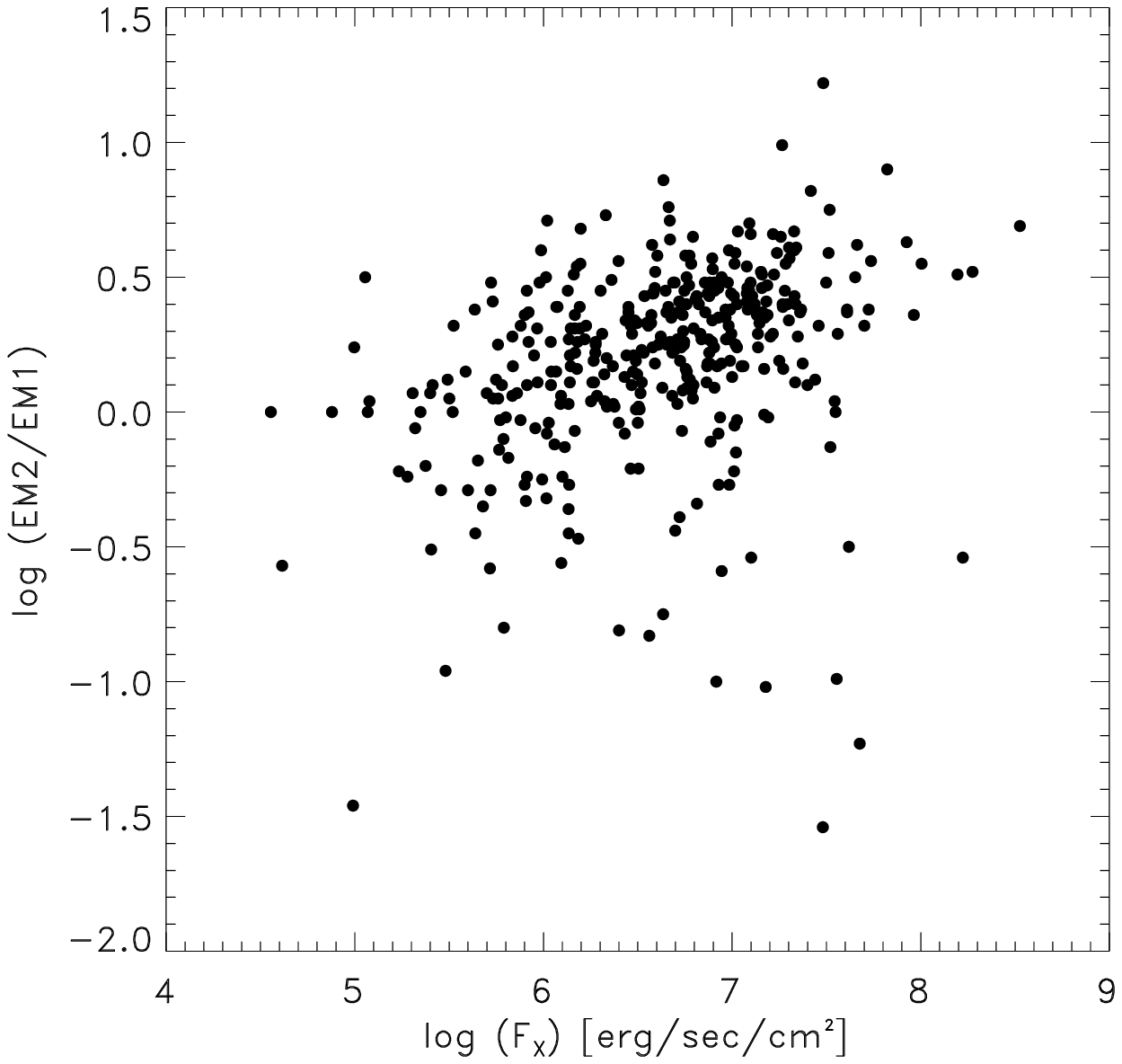}
\figcaption{ Left: Plasma temperatures (crosses for $T_{\rm X1}$, diamonds
for $T_{\rm X2}$) derived in the X-ray spectral fits
for the TTS in the COUP optical sample
plotted versus the X-ray surface flux. 
The dashed line shows the
relation $F_X \propto T^6$.\newline
Right: Emission measure ratio of hot and cool plasma component
versus the X-ray surface flux.
\label{tx_fx.fig}}
\end{figure}

In Fig.~\ref{tx_fx.fig} we plot the plasma temperatures 
and the ratios of the emission measures of the hot and cool component
derived in the spectral fits versus the X-ray surface flux.
First, we  note that the temperature of the hot plasma component 
increases with increasing surface flux; the slope 
is consistent the relation $F_X \propto T^6$, what is 
similar to a scaling relation derived for the
solar corona by \cite{Peres04} (which however, was established for a much
lower temperature range than seen here on the TTS).
Second, we note that the relative contribution from the
hot component (as measured by the ratio of emission measures for the
hot and cool components) also increases with increasing X-ray activity.
The most interesting result from these plots is
the remarkable similarity of the temperatures of the cool plasma 
component in our sample. For nearly all stars a temperature of about 10~MK
is found for the cool component. This suggests that 
this 10~MK component is a real feature 
in the coronal temperature
distribution of the TTS. 
It is interesting to note that a $\sim 10$~MK plasma component seems
to be some kind of a general feature of coronally active stars; for example,
\cite{Sanz03} determined the emission measure distribution of
28 coronally active nearby field stars 
and found a peak at 8--10~MK in most of their stars.
They argued that this temperature component may define a fundamental coronal
structure, which is probably related to a class of compact loops
with high plasma density.

\begin{figure}[p]
\plotone{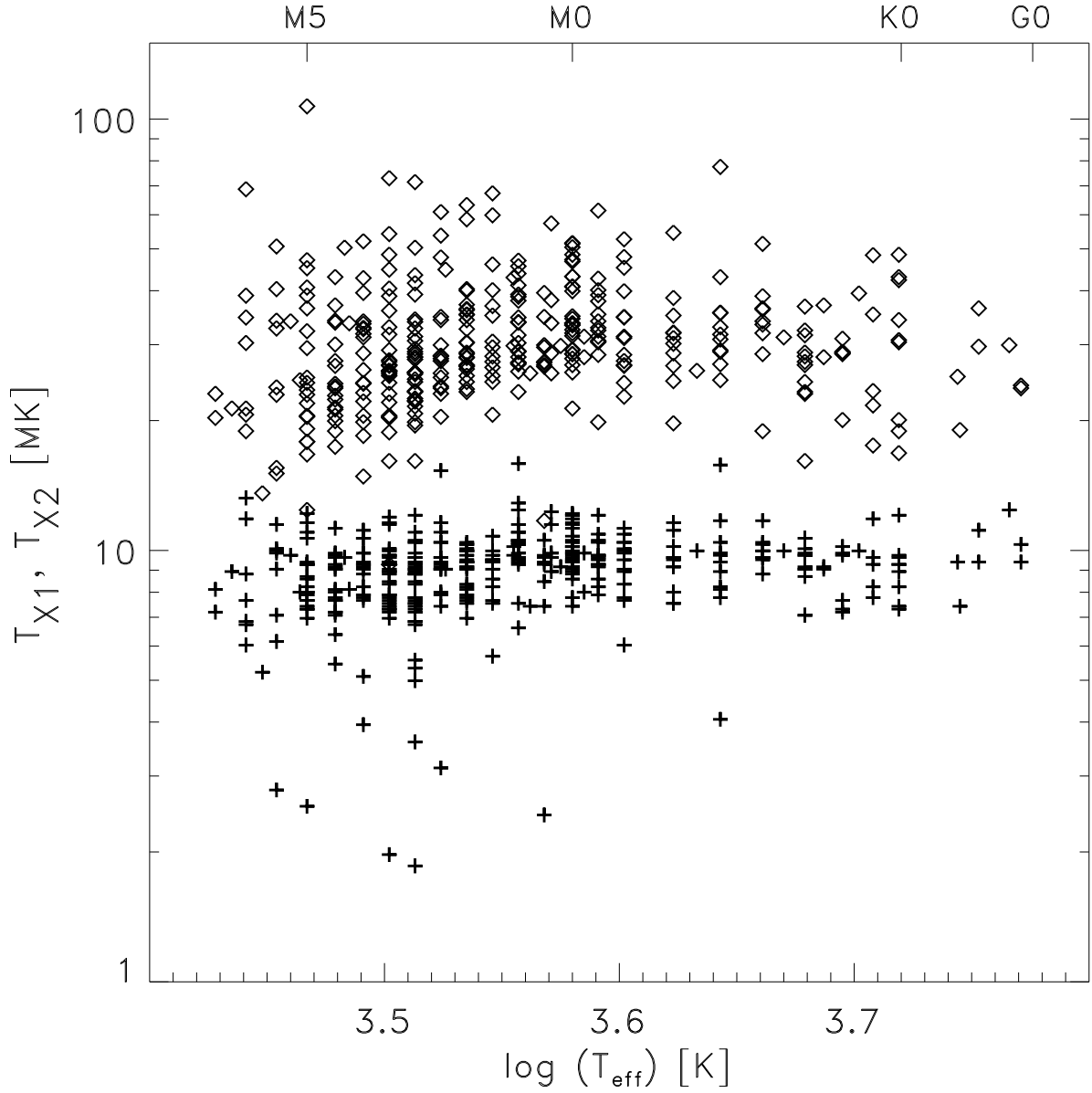}
\figcaption{Plasma temperatures (crosses for $T_{\rm X1}$, diamonds
for $T_{\rm X2}$) derived in the X-ray spectral fits
for the TTS in the COUP optical sample
plotted versus the effective temperature.
\label{tx_teff.fig}}
\end{figure}

Figure \ref{tx_teff.fig} shows  the derived plasma temperatures
as a function of stellar effective temperature.
Among the M-type stars one can see a decrease in the temperature of the
hot component for decreasing effective temperature. This can be understood
as a consequence of  the correlation of the hot plasma temperature
with the level of X-ray activity (as traced by $F_{\rm X}$ or
$L_{\rm X}/L_{\rm bol}$) and the decrease of X-ray activity with decreasing
mass or effective temperature (see, e.g.~Fig.~\ref{fx_teff.fig}).

\begin{figure}[p]
\plotone{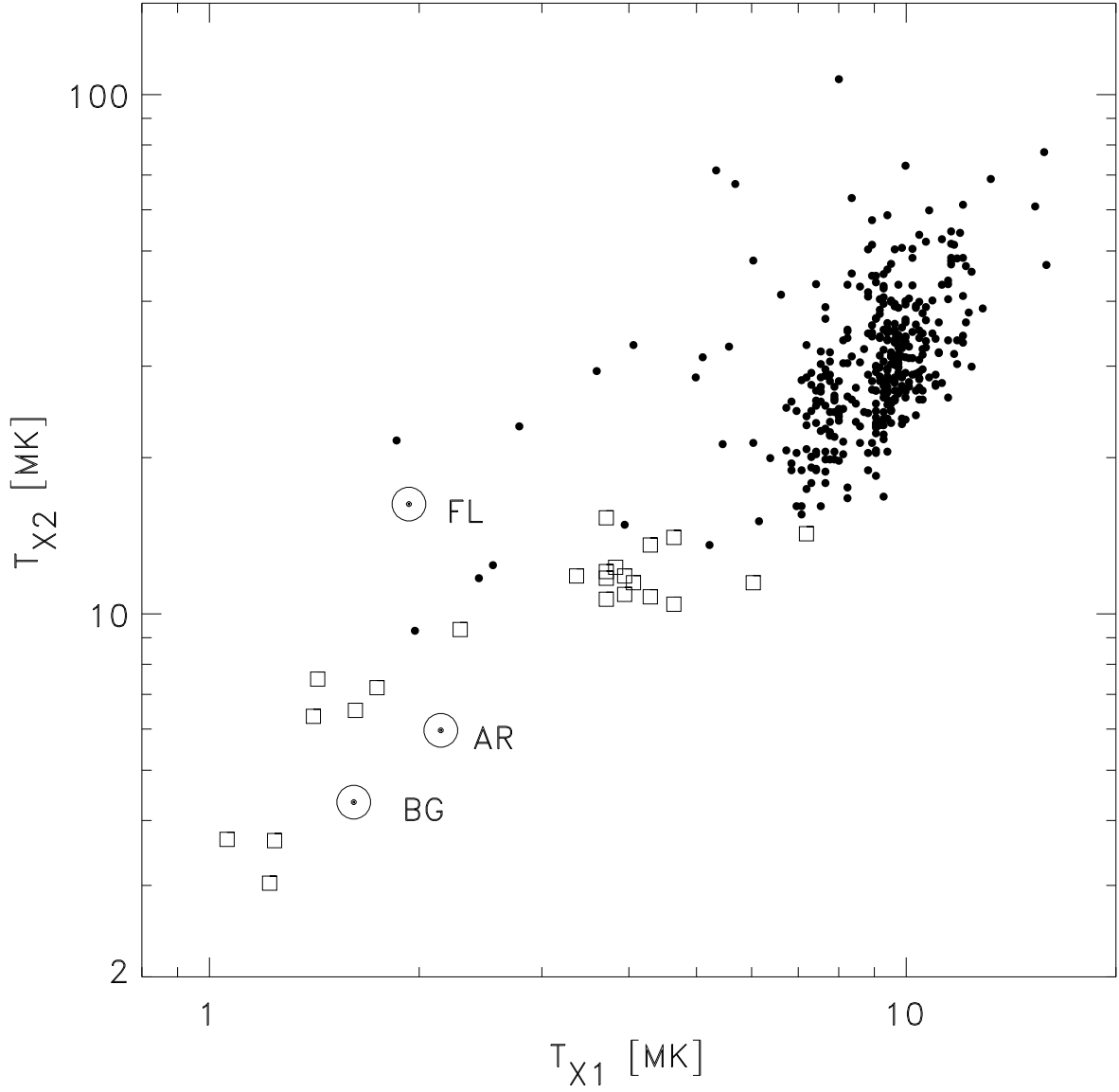}
\figcaption{Temperature of the hot versus the cool
plasma component for the TTS in the COUP optical sample (solid dots).
The open squares show plasma temperatures derived for G- and K-type MS stars
\citep{Briggs03, Pillitteri04, Guedel97},
and typical values for structures in the solar corona
are also shown (BG = background corona, AR = active region, FL = flare;
from \citealp{Orlando04}).
\label{tx1_tx2.fig}}
\end{figure}

In Fig.~\ref{tx1_tx2.fig} we plot the temperature of the hot versus that of the
cool plasma component for the ONC TTS.
We also have included temperatures derived for
G- and K-type stars in several young clusters and for solar-like
field stars,
as well as values for different structures in the solar corona
as derived in the simulations of the ``Sun as a star" by \cite{Orlando04}.
The ONC TTS follow the general
correlation between the temperatures of
the hot and cool plasma components seen for the MS stars,
although nearly all TTS show much higher temperatures than found on the
MS stars.
In the solar corona and in many active MS stars, 
plasma temperatures significantly exceeding 10~MK are only seen during
strong flares. The high plasma temperatures found for the TTS may 
suggest an increased contribution of flares to the total X-ray emission.

\section{X-ray emission and circumstellar accretion disks\label{accretion.sec}}

It is still unclear how the (magnetic) interactions between a 
young stellar object and its surrounding circumstellar accretion disk
influence the X-ray activity.
Many X-ray observations of star-forming regions have been used to 
search for differences in the X-ray properties of the classical T Tauri stars
(CTTS; usually defined by the presence of H$\alpha$ emission with
equivalent widths $W({\rm H}\alpha) \ge 
10$ \AA), which are thought to be actively accreting via circumstellar 
disks, and the weak-line T Tauri stars 
(WTTS; $W({\rm H}\alpha) < 10$ \AA), which seem
to lack disks and active accretion. 
The results of these investigations showed confusing differences:
some studies (e.g.~Gagn\'e et al.~1995; Feigelson et al.~1993;
Casanova et al.~1995; Preibisch \& Zinnecker 2002) found  
no significant differences 
between the X-ray luminosity functions of CTTS and WTTS,
while other studies, most notably the Taurus-Auriga study by
\cite{Stelzer01}, found clear differences 
in the X-ray luminosity functions,
with the WTTS being the stronger X-ray emitters.
The two $Chandra$ studies of the ONC before COUP also yielded seemingly
contradictory results: \cite{Feigelson02a} found no differences in the
X-ray activity levels of stars with and without disks, whereas
\cite{Flaccomio03b} reported a strong difference in the X-ray activity levels of
accreting and non-accreting stars.

It is important to note that the different studies used different
criteria to define the samples of CTTS and WTTS: sometimes 
the strength of the H$\alpha$ or Ca~emission line were used, while other
studies used 
the presence or absence of near-infrared excess emission.
These different kinds of indicators actually measure different things that cannot
be directly compared: H$\alpha$ or Ca~line emission is thought to be
a tracer of accretion, while infrared excess emission is a tracer of
circumstellar material. While accretion obviously requires the presence
of circumstellar material, the presence of circumstellar material alone
does not necessarily mean that the young stellar object is also accreting.
Furthermore, this issue is easily affected by strong selection
effects if the samples of TTS are either X-ray selected or based
on optical selection criteria such as emission lines or infrared excess. 
For example, in many star forming regions H$\alpha$ emission was used 
as a tracer to find and define the population of T Tauri stars. 
This can easily introduce a strong bias,
because the CTTS are quite easy to recognize by their prominent 
H$\alpha$ emission
even if they are very faint, whereas WTTS of similar brightness
are much harder to identify. The optical H$\alpha$ selected samples
of TTS are therefore often very incomplete for WTTS and much more complete
for the CTTS \citep[see discussion in][]{PZ02}. 
In fact, the majority of WTTS in many
star forming regions have been found through X-ray observations (cf.~Neuh\"auser
1997) and therefore suffer from an X-ray selection bias.

The COUP study provides us with the important advantage that we can
use a statistically complete sample of all optically visible stars 
in the region, which does not suffer from any of the selection effects
described above. In the following we will investigate how and in which 
way the X-ray activity is related to infrared excess emission as a
tracer of circumstellar material (\S\ref{lx_ex}), optical line
emission as a tracer of accretion (\S\ref{lx_ca}), and estimates
of accretion rates and luminosities from astrophysical
models (\S\ref{lx_acc}).

\subsection{X-ray activity and infrared excess (= inner disk tracer)\label{lx_ex}}

A good way to discern between TTS with and without circumstellar
material is to look for infrared excess emission.
The COUP tables list the color excess $\Delta\left(I-K\right)$ as
determined by \cite{Hillenbrand98}. 
This quantity represents the color excess relative
to the expected photospheric colors for the star's spectral type
after correction for reddening due to extinction, and
 has been shown to be a useful tracer of 
circumstellar material \citep[see discussion in][]{Hillenbrand98}. However, the $\Delta\left(I-K\right)$ excess
is not optimal for detecting circumstellar material, since the
$K$-band excess traces only the hottest dust in the 
innermost regions near the central star; it has been shown 
(e.g.~Haisch et al.~2001) that 
many stars with circumstellar material show significant excess emission
only at longer wavelengths.  We therefore also used the $L$-band photometric
data listed in the COUP tables and determined a color excess 
$\Delta\left(K-L\right)$ in an analog way as $\Delta\left(I-K\right)$.
The $\Delta\left(K-L\right)$ excess has the advantage of being more
sensitive for circumstellar material and less strongly affected by 
uncertainties in the visual extinction, but the disadvantage 
of being available for fewer (228) TTS than $\Delta\left(I-K\right)$ 
(446 TTS) in the optical sample.

\begin{figure}[p]
\plotone{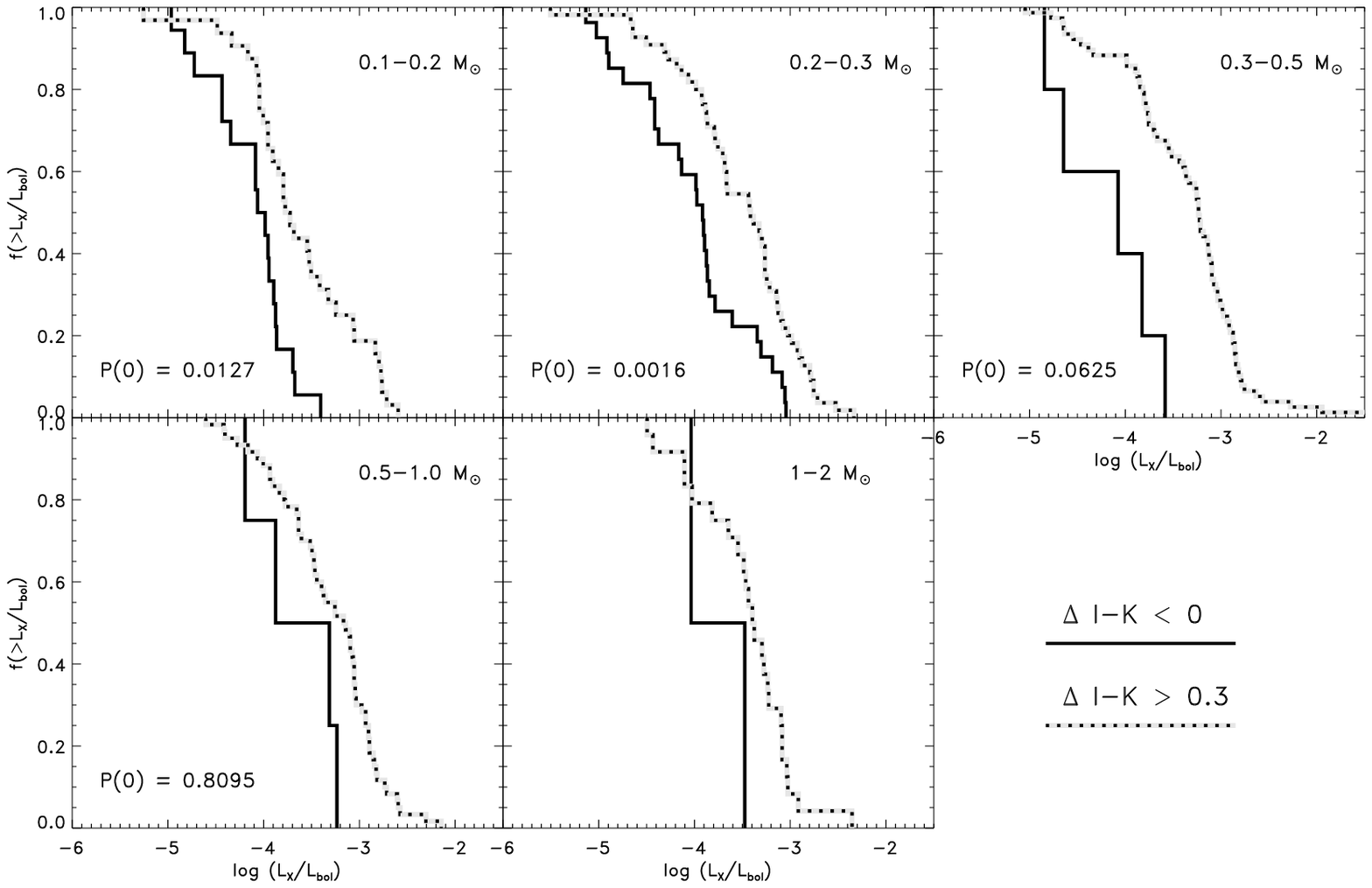}
\figcaption{Cumulative distributions of the fractional X-ray luminosities
for TTS with ($\Delta\left(I-K\right) >0.3$) and without
($\Delta\left(I-K\right) <0$) infrared excess
in the lightly absorbed optical sample
for five different mass
ranges. The KS test probabilities for the assumption that both samples
are drawn from the same underlying distribution are given in the
lower left edge of each plot.
\label{lxlb_delik.fig}}
\end{figure}

\begin{figure}[p]
\plotone{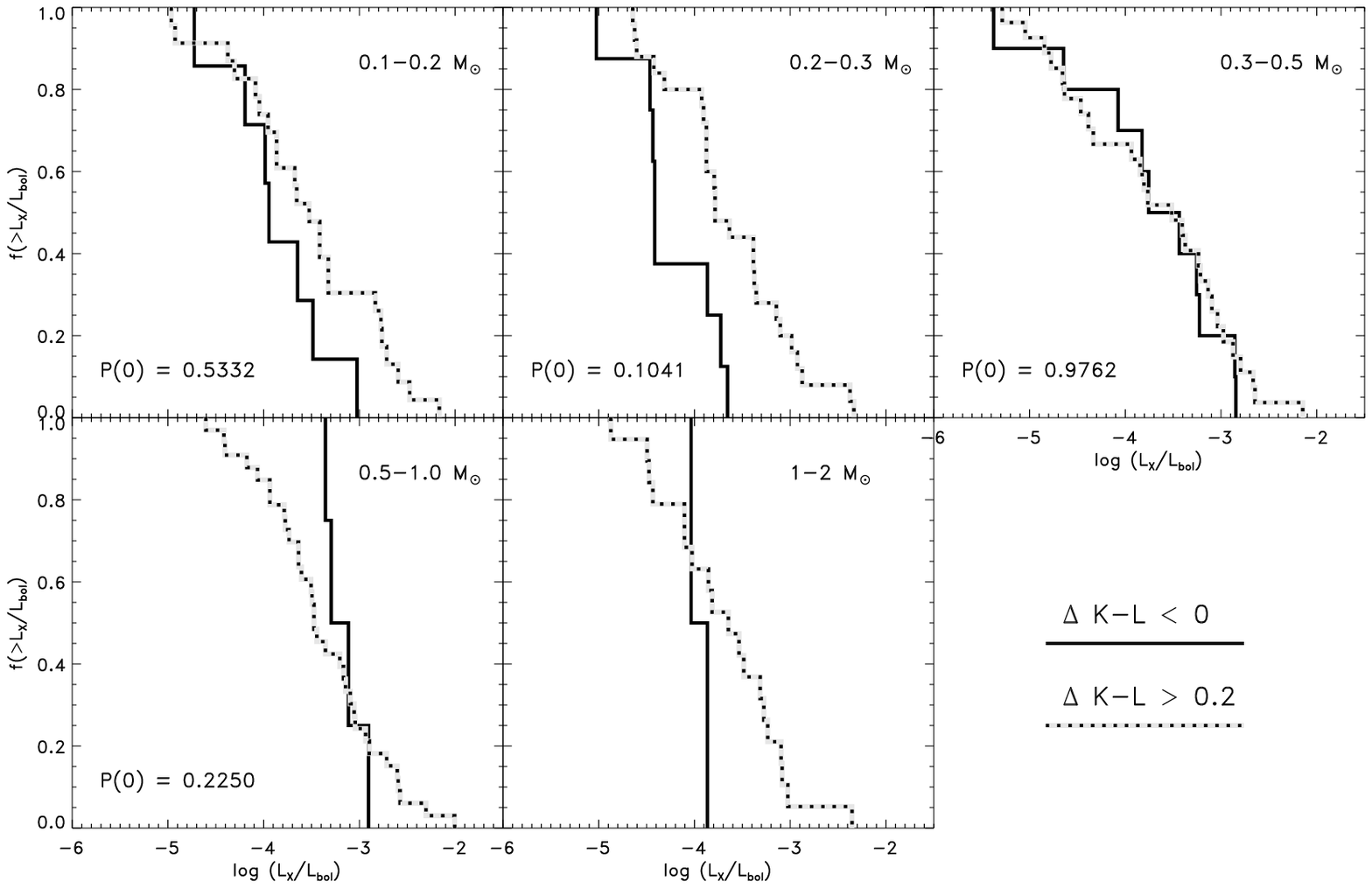}
\figcaption{
Cumulative distributions of the fractional X-ray luminosities
for TTS with ($\Delta\left(K-L\right) >0.2$) and without
($\Delta\left(K-L\right) <0$) infrared excess
in the lightly absorbed optical sample
for five different mass
ranges. The KS test probabilities for the assumption that both samples
are drawn from the same underlying distribution are given in the
lower left edge of each plot.
\label{lxlb_delkl.fig}}
\end{figure}

Since the
infrared excess emission in our sample is correlated to 
effective temperature and stellar mass, we compare 
in Fig.\ref{lxlb_delik.fig} and Fig.~\ref{lxlb_delkl.fig} stars with and
without excesses in mass stratified samples.
In the samples based on $\Delta(I-K)$ excess, 
we find significantly different ($P(0) < 0.02$)
fractional X-ray luminosity distributions for the $0.1-0.2\,M_\odot$ and
the $0.2-0.3\,M_\odot$ bins, whereas for the more massive stars
the differences are only of marginal statistical significance.
If we use the $\Delta(K-L)$ excess (which we regard as more reliable),
no significant differences are found for any of the mass ranges
considered.

%%%%%%%%%%%%%%%%%%%%%%%%%%%%%%%%%%%%%%%%%%%%%%%%%%%%%%%%%%%%%%%%%%%%%%%%%%%%

\subsection{X-ray activity and Ca~II line emission (= accretion tracer)\label{lx_ca}}

Next we consider the presence/absence of signs for
active accretion rather than disks. We follow the strategy of
\cite{Hillenbrand98}, who used the equivalent width of the $8542$\,\AA\,
Ca II line as an indicator of disk accretion. As they noted,
stars without or with very weak accretion are expected to show this line 
in absorption
with $W({\rm Ca~II}) \sim 3$\,\AA\, and a rather weak dependence on the spectral type.
In more strongly accreting stars, the line gets filled and the equivalent 
width should be related to the mass accretion rate.
In their analysis of the {\it Chandra} HRC data of the ONC,
Flaccomio et al.~(2003) classified stars as strong accretors
if their Ca~II line is in emission and has an equivalent width of
 $W({\rm Ca~II}) < -1$\,\AA, while stars with the Ca II line in absorption
with $W({\rm Ca~II}) > 1$\,\AA\, are
assumed to be not (or at most very weakly) accreting.
Using this scheme, 142 (136) of the objects in our (lightly absorbed)
optical sample are classified as strong accretors, and 134 (123) objects
as weak or non-accretors. In the following text, we will simply denote these
two groups as ``accretors'' and ``non-accretors''.

\begin{figure}[p]
\plotone{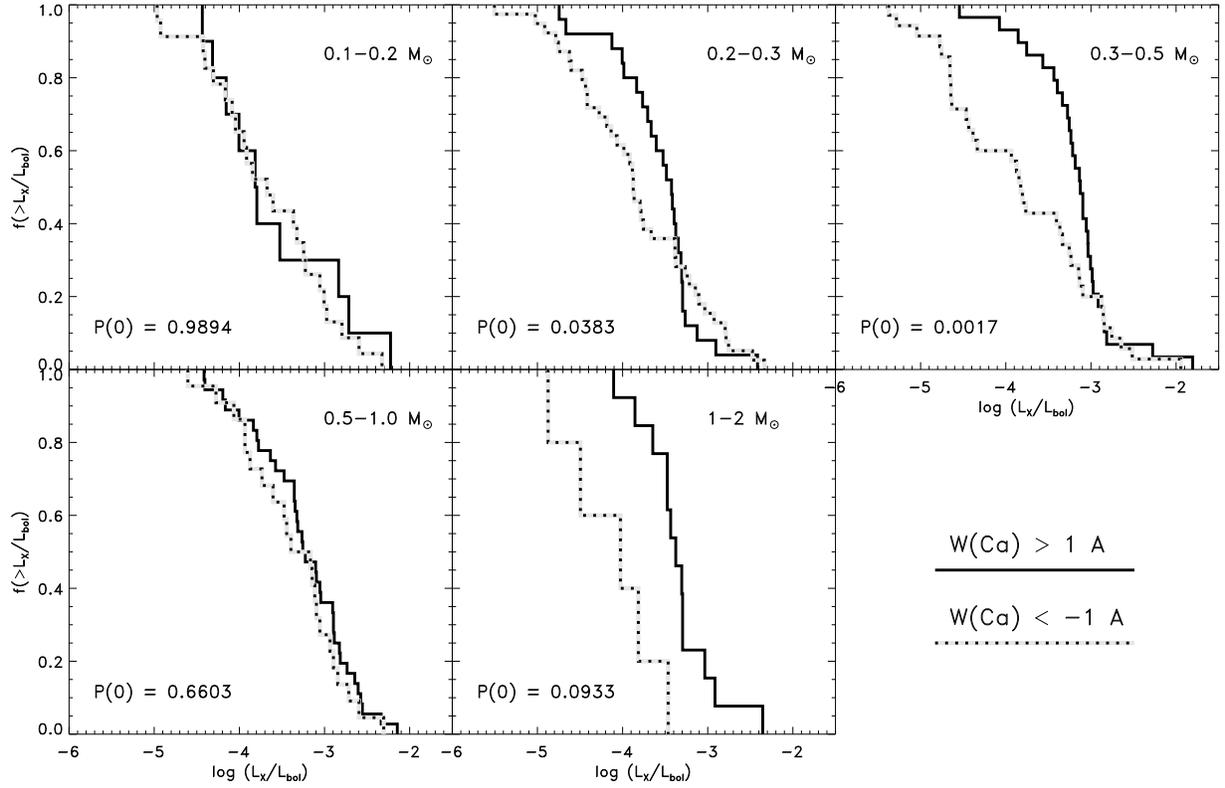}
\figcaption{Cumulative distributions of the fractional X-ray luminosities
for accreting and non-accreting TTS
in the lightly absorbed optical sample
for five different mass
ranges. The KS test probabilities for the assumption that both samples
are drawn from the same underlying distribution are given in the
lower left edge of each plot.
\label{lxlb_ewca.fig}}
\end{figure}

Flaccomio et al.~(2003) found a clear difference in the (fractional)
X-ray luminosities of the accretors and non-accretors in 
their $Chandra$ HRC data of the ONC, with the accretors being
considerably (about a factor of 2--3) less X-ray bright than
the non-accretors. A similar, although less pronounced, effect is found
in our COUP data. In Fig.~\ref{lxlb_ewca.fig} we show the distributions
of fractional X-ray luminosities of accreting and non-accreting stars
in different mass bins.
A significant difference between accretors and non-accretors is only
found for the $0.3-0.5\,M_\odot$ stars, whereas the other mass 
ranges show only marginal or no evidence at all for a difference in the
distribution functions of fractional X-ray luminosities.

\begin{figure}[p]
\plottwo{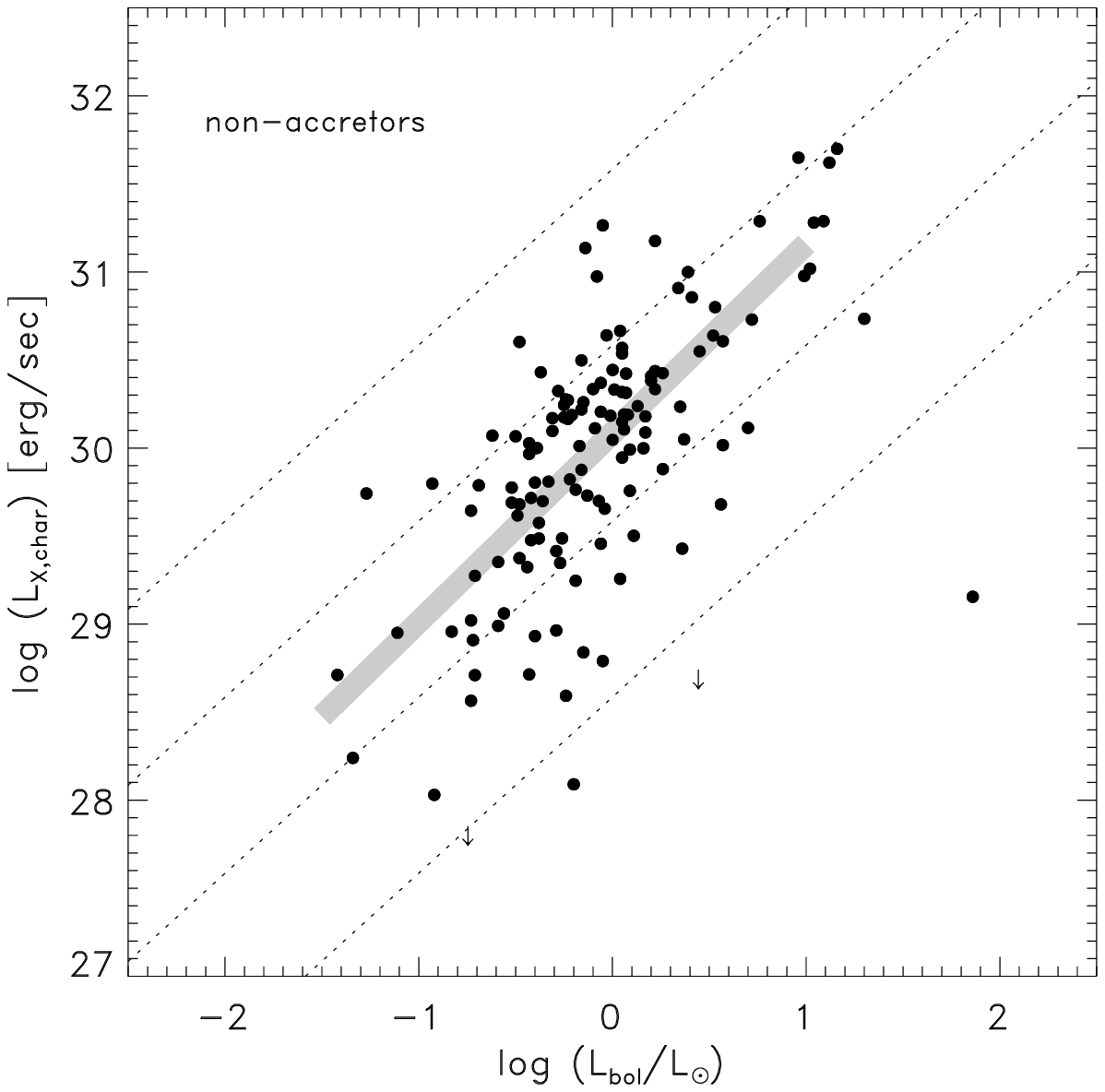}{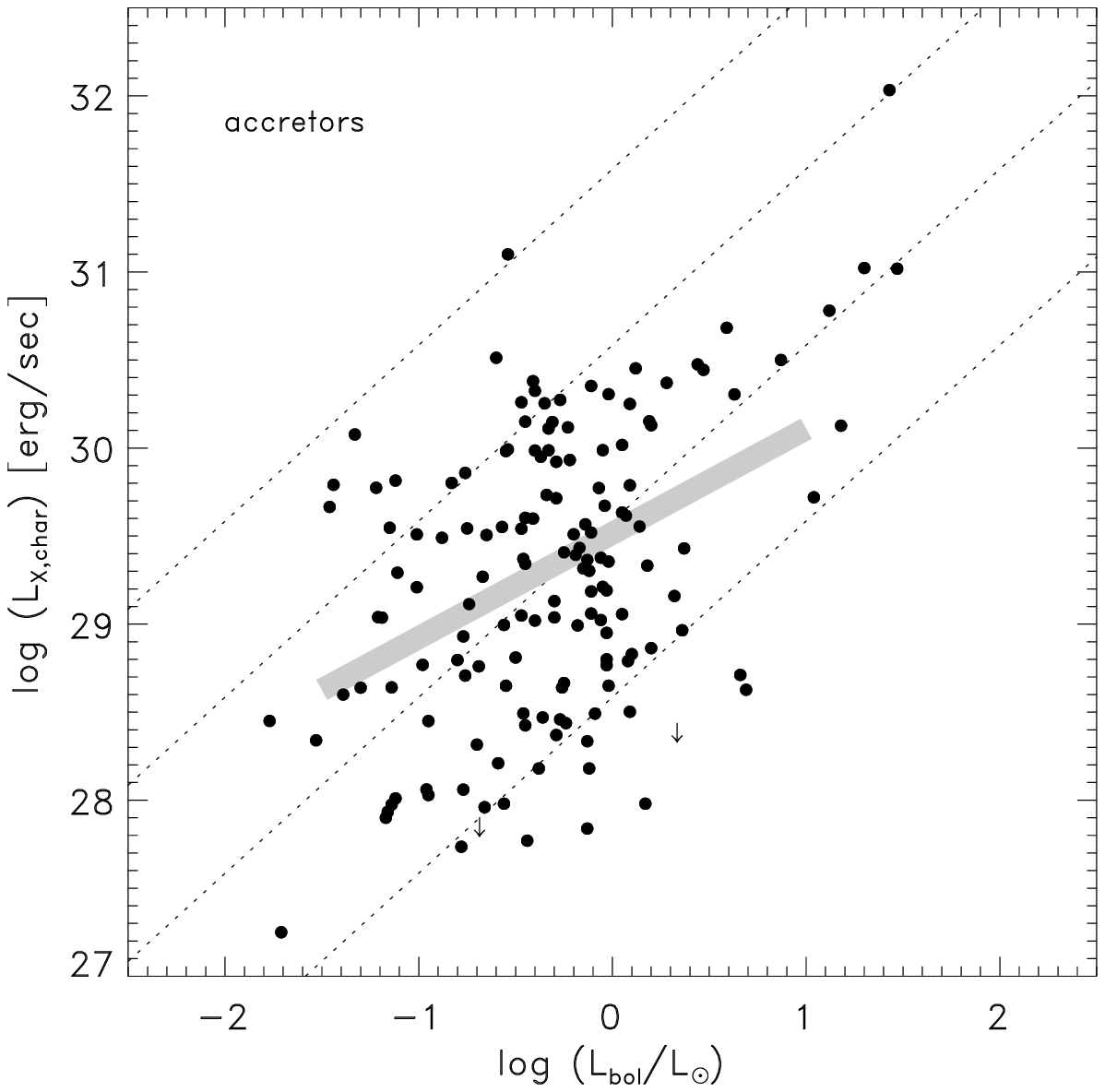}
\figcaption{Characteristic X-ray luminosity versus
bolometric luminosity for the stars
in the optical sample with the $8542$\,\AA\,Ca~II line in absorption (left,
non-accretors) and in emission (right, accretors).
The dotted lines mark $\log\left(L_{\rm X}/L_{\rm bol}\right)$ ratios of  $-2$,
$-3$, $-4$, and $-5$. The thick grey lines
shows linear regression fits for $L_{\rm bol} \leq 10\,L_\odot$ stars
 with the EM algorithm computed with ASURV.
\label{lx_lb_ca.fig}}
\end{figure}

To investigate the difference between accreting and non-accreting stars
further, we compare in Fig.~\ref{lx_lb_ca.fig} the correlations between
characteristic X-ray luminosities and the bolometric luminosities for
the non-accretors and the accretors.
The non-accretors
show a very well defined correlation between $L_{\rm X,char}$ and 
$L_{\rm bol}$, the
linear regression fit with the EM algorithm gives a 
power-law slope $1.1\pm 0.1$ and standard deviation of
$0.52$~dex in $\log\left(L_{\rm X,char}\right)$. 
For the accretors, the correlation is much less well defined;
the linear regression fit with the EM algorithm gives a
power-law slope $0.6\pm 0.1$ and standard deviation of
$0.72$~dex in $\log\left(L_{\rm X,char}\right)$.
Very similar results are found for the correlations between
characteristic X-ray luminosities and stellar masses for
the non-accretors and the accretors.

As discussed in \S\ref{variability.sec}, the scatter in the correlations
due to X-ray variability, the uncertainties in the variables, and
unresolved binaries is expected to be about $0.4\!-\!0.5$~dex.
The standard deviation found for the $L_{\rm X}\leftrightarrow L_{\rm bol}$
correlation (or the  $L_{\rm X}\leftrightarrow M$ correlation) for the
non-accreting stars agrees well to that expectation, whereas the
accreting stars show a considerably larger scatter.
Some fraction of this scatter may be due to the fact that
the more rapidly accreting stars may have larger errors compared to
non-accreting stars in stellar luminosity and effective temperature 
values due to the effects
of accretion on the observables that lead to these quantities.

Another important result is found when considering the mean fractional
X-ray luminosities: For the non-accretors, 
the median value for  $\log\left(L_{\rm X}/L_{\rm bol}\right)$
is $-3.31$, what is in very good agreement 
to the mean saturation value for rapidly rotating low-mass 
($0.22 - 0.6\,M_\odot$) field stars derived by \cite{Pizzolato03}.
This means that the fractional X-ray luminosities of the 
non-accreting TTS 
are consistent to those of much older coronally active field stars.
For the accretors, on the other hand, we find a
median value for  $\log\left(L_{\rm X}/L_{\rm bol}\right)$
of $-3.74$, which is a factor of about 3 lower than the 
saturation value for fast rotating low-mass 
field stars. The accreting TTS therefore are responsible
for the ``X-ray deficit" of the ONC TTS, i.e.~the fact that the median
fractional X-ray luminosity of the TTS is lower than that 
found for rapidly rotating MS stars.

%%%%%%%%%%%%%%%%%%%%%%%%%%%%%%%%%%%%%%%%%%%%%%%%%%%%%

We also
looked for possible relations between the X-ray activity and the
rotation rates of the accretors and non-accretors.
The rotation periods found for the non-accretors and accretors
overlap strongly, but the non-accretors show
shorter median periods ($3.8$ days)
than the accretors ($6.8$ days); a KS test gives
a probability of only 0.03 that the distribution
of periods is identical in both groups.
However, neither for the accretors nor for the non-accretors statistically
significant correlations are found between activity and rotation.
The median fractional X-ray luminosities of the non-accretors  and accretors
with known rotation period are nearly identical,
$-3.3\pm0.4$ and $-3.4\pm 0.8$.
The difference in X-ray activity between accreting and non-accreting stars
described above
can thus not be explained by differences in their rotation periods
or their rotation-activity relations.
Considering the relation between X-ray activity and Rossby numbers,
we also find no significant correlations for either accretors or non-accretors.

\subsection{X-ray activity and accretion rates/luminosities\label{lx_acc}}

\cite{Robberto04} recently determined
accretion rates and accretion luminosities for a sample of 40
TTS in the Trapezium cluster from HST $U$- and $B$-band photometry.
As the computation of the accretion parameters                          
from the UV excess is quite
indirect and involves numerous assumptions, their values have
considerable uncertainties; for some of their
stars they find even negative values for the accretion luminosities.
We therefore restrict us to those 30 stars
for which they derived positive values for the accretion luminosity
and note that 29 of these are detected as X-ray sources in COUP.

For a considerable fraction of these objects the
X-ray luminosities are comparable or even larger than the accretion 
luminosities;
this is a strong argument against the assumption that
that the X-ray emission in these TTS is directly created in the accretion 
process (e.g., comes from the accretion shock; see discussion below).
Furthermore, we find a weak anti-correlation of the fractional 
X-ray luminosity with accretion rate (and also with accretion luminosity); 
although
these correlations are not statistically significant, 
they agree to the 
above results based on the Ca line width classification and suggest
that active accretion somehow lowers the X-ray activity levels.

\subsection{Summary of the relations between X-ray activity and 
accretion}

We find that the TTS with inner circumstellar material as traced
by near-infrared excess show slightly higher fractional X-ray
luminosities than the TTS without near-infrared excess,
but this difference is of only marginal statistical significance.
Using the equivalent width of the Ca~II line to discern between 
accreting and non-accreting stars, we find that 
the non-accretors show very well defined correlations
between X-ray luminosity and bolometric luminosity or stellar mass.
The accreting stars, on the other hand, produce much poorer correlations
between $L_{\rm X}$ and $L_{\rm bol}$ or stellar mass and much more scatter.
Furthermore, the mean fractional X-ray luminosities of the non-accreting
TTS are well consistent with those of rapidly rotating MS stars,
while the accreting TTS show about 3 times lower levels of X-ray activity.
Finally, we have shown that X-ray activity appears to be anti-correlated
with mass accretion rate.

In conclusion, the X-ray activity of the non-accreting TTS is consistent
with that of rapidly rotating MS stars,
while in accreting TTS  the X-ray emission is somehow suppressed.

\section{Implications on the origin of TTS X-ray emission\label{conclusions.sec}}
In this section we summarize what implications we can derive from our 
X-ray data on the origin of the observed X-ray emission from the TTS.
We will consider the following questions:
Where is the X-ray emitting plasma located?
What is the reason for the lower X-ray activity of accreting stars
in comparison to non-accretors?
What is the ultimate origin of the magnetic activity and what 
kind of dynamo may work in the TTS?

\subsection{Location of the X-ray emitting structures}

\subsubsection{X-ray emission from accretion shocks?}

According to the magnetospheric accretion
scenario, accreted material crashes onto the stellar surface with
velocities of up to several 100 km/sec, what should cause hot 
($\lesssim 10^6$~K)
shocks in which strong optical and UV excess emission
and perhaps also soft X-ray emission is produced 
\citep[e.g.,][]{Lamzin96, Calvet98}.
The expected characteristics of
X-ray emission from accretion shocks would be a very soft spectrum
(due to the low plasma temperature in the shock of at most a few MK),
and perhaps simultaneous brightness variations at optical/UV 
wavelengths and in the X-ray band. Although earlier studies failed to
find evidence for this scenario \citep[e.g.,][]{Gullbring97},
more recent high-resolution X-ray spectroscopy of {\em some}
TTS  \citep[e.g.~TW Hya, XZ Tau
and BP Tau, see][]{Kastner02, StelzerSchmitt04, Favata03, Schmitt05}
yielded plasma temperatures and densities that
have been interpreted as evidence for X-ray emission originating from
a hot accretion shock.

The COUP results provide no support for a scenario in which X-ray
emission is dominated by accretion power. First, 
we note that many of the accreting TTS show X-ray
luminosities considerably larger than their total accretion
luminosities. Although we have to be somewhat cautious because the
X-ray and accretion rate measurements were not simultaneous and
accretion is thought to be strongly time variable, it appears
extremely unlikely that the bulk of the observed X-ray emission from
the TTS could originate from accretion processes. 
Second, the COUP spectra of nearly all TTS show much higher plasma
temperatures (typically a $\sim 10$~MK cool component and $\gtrsim
20$~MK hot component) than the $\lesssim 1-3$~MK expected from shocks
for the typical accretion infall velocities. 
Only for five of the TTS in our optical sample the X-ray spectral fits
yielded plasma components with temperatures below 3~MK.
For none of these stars accretion rates are known,
only two of them show Ca~II emission, and
only one displays infrared excess. Thus, we cannot determine
whether in any of these objects the X-ray emission may be related to
accretion shocks.
We just note that the low plasma temperatures alone
provide no evidence for an accretion shock origin of the X-rays,
since similarly low (or even lower) plasma temperatures are found
for MS stars and the Sun, i.e.~for stars that are certainly not accreting.
Furthermore, the COUP
lightcurves show thousands of high-amplitude flares whose temporal
structure closely resembles solar-type magnetic flares,
and is very unlikely to be reproduced by thermal accretion shocks.
Indeed, \cite{Stassun05} have compared
these COUP lightcurves with simultaneous optical lightcurves and find
very little evidence to suggest that X-ray variability is correlated
with accretion-related processes.

Of course, these arguments do not exclude the possibility 
that accretion shocks may contribute {\em some fraction}
 of the X-ray emission in TTS. 
It is critical to note that $Chandra$'s ACIS-I instrument is
not very sensitive to the cooler plasma expected from these
accretion shocks, and much of this emission may be attenuated
by line-of-sight interstellar material.
We note that
evidence for a scenario of mixed X-ray emission in
accreting YSOs has recently come from an X-ray high spectral
resolution observation of BP~Tau \citep{Schmitt05}, where 
hot plasma (too hot to be shock-produced and thus likely
magnetically confined and heated in some form of coronal structure)
coexists with cool (1--3~MK) plasma for which unusually high densities were
inferred, which may be well explained by accretion shock models\footnote{The 
cool plasma in BP~Tau shows a very low value of the ratio between
the intensity of the forbidden and intercombination lines ($R = f/i$)
for the O\,{\sc vii} triplet, formed at temperatures of 1--3 MK. Low
$R$ values (also observed in TW Hya, and in no coronal source) can
either imply very high densities, or the presence of an intense UV
field, as indeed expected within the accretion spot. Too high
densities would however be difficult to explain given the pressure
structure of an accretion shock, (see \citealp{Drake05}), in which the
plasma at $n_e \simeq 10^{13}$ cm$^{-3}$ (the density implied by the
$R$ value observed in TW~Hya) is buried (following e.g.\ the shock
structure and emission models by Calvet \& Gullbring 1998) under a
column density of typically $\gtrsim 10^{22}\,{\rm cm}^{-3}$, that
should absorb the soft X-ray emission from the shock zone nearly
completely and thermalize the high-energy radiation within or close to
the shock.}.

We conclude that, although accretion shock emission must be
present, plays an important role in optical or ultraviolet
emission of CTT stars \citep[e.g.][]{Lamzin96}, and may contribute
some fraction of the largely-unobserved soft X-ray emission of TTS, it is not
responsible for the bulk of the X-ray emission seen in the COUP data.

\subsubsection{X-ray emission from magnetic star-disk interactions?}

Another possibility for a non-solar like origin of the X-ray radiation
may be plasma trapped in magnetic fields that connect the star with
its surrounding accretion disk.  The dipolar stellar magnetic field
lines anchored to the inner part of the accretion disk should be
twisted around because of the differential rotation between the star
of the disk. Theoretical work suggests that the twisted field lines
periodically reconnect, and the released magnetic energy heats plasma
trapped in the field lines to very hot, X-ray emitting temperatures
(Hayashi et al.~1996; Montmerle et al.~2000; \citealp{Iobe03}; 
\citealp{Romanova04}).

The analysis of the $\sim 30$ largest flares in COUP data
by \cite{Favata05} suggests
that very long magnetic structures (up to a few times $10^{12}$ cm)
are present in some of the most active stars in our sample. Such large
structures (tens of times the size of the star) may indicate a magnetic
link between these stars and their disks.
However, we note that very large loop lengths 
were derived for only a few of these flares; for the majority
of the analyzed flares much smaller loop lengths were found.

Furthermore, our results in the previous sections 
show that, in general, the X-ray luminosity is strongly linked
to stellar parameters like bolometric luminosity and mass, but does
not strongly depend on the presence or absence of circumstellar disks
as traced by near-infrared excess emission. 
It is reasonable to conclude that although X-ray
emission from magnetic star-disk interactions may be seen
during some of the most intense flares, the bulk of the
observed X-ray emission from ONC TTS probably originates
from more compact structures with geometries resembling
solar coronal fields.

\subsubsection{X-ray emission from solar-like coronal structures?}

Our data are generally consistent with the assumption that the
observed X-ray emission originates from, in principle solar-like,
coronal structures. The X-ray luminosities and plasma temperatures
derived for the ONC TTS are comparable to those of the most
active MS stars, and can thus be understood by assuming
that stellar coronae in general are composed of X-ray emitting structures
similar to those present in the solar corona (e.g.~Drake et al.~2000;
Peres et al.~2004). 
Although the high X-ray luminosities of active MS stars
and TTS can not be reproduced by simply filling the available 
coronal volume with solar-like active regions,
coronal structures with higher plasma density\footnote{Remember that the
emission measure, $EM :=\int  n^2\, {\rm d}V$, scales linearly with the
plasma volume $V$, but with the
square of the density $n$.}  can explain the
high levels of stellar X-ray activity. 
Evidence for higher than solar plasma densities is found in
high-resolution X-ray and EUV spectra for many active stars
\citep[e.g.][]{Sanz03, Ness04}.
Furthermore, once the stellar coronae get nearly completely filled
with active regions, the magnetic interaction of the active regions
should lead to increased flaring in the most active stars, boosting
their X-ray luminosities even further (Peres et al.~2004).

We also note that the temporal behavior of most flares
seen in the COUP data is rather similar to what is seen in flares
on the Sun or active MS stars  
(Wolk et al.~2005). Further evidence suggesting enhanced
solar-like coronal activity as the source of the
X-ray emission from active MS stars and TTS is summarized 
e.g., in Feigelson \& Montmerle (1999) and Favata \& Micela (2003).

\subsection{The suppression of X-ray emission by accretion}

Our COUP data confirm previous results that accreting TTS 
show lower levels of X-ray activity than non-accretors
(Stelzer \& Neuh\"auser 2001; Flaccomio et al.~2003; Stassun et al.~2004a).
Here we discuss some possible explanations for this effect.
Two previously suggested explanations can essentially be ruled out
by our data. The first one is the suggestion that the systematically 
higher extinction of the
accreting CTTS (due to absorption in the accretion disk)
may be responsible for their weaker {\em observed} X-ray emission
as compared to the WTTS \citep[e.g.][]{Stassun04a}. In our COUP data,
individual extinction-corrected X-ray luminosities could be determined
in a self-consistent fitting analysis of the individual X-ray spectra.
The different levels of extinction in the
accreting and non-accreting stars should not lead to errors in the
extinction-corrected X-ray luminosities.

We can also exclude the idea that accreting TTS are weaker X-ray emitter 
because their rotation is braked by magnetic disk locking, leading  
to weaker dynamo action and therefore less X-ray emission than
in the non-braked WTTS. We have shown in \S\ref{lxlb_prot.sec} that 
neither the accretors nor the non-accretors show a relation 
between rotation and X-ray activity, and thus the difference in 
rotation rates cannot be the reason for the
difference in the level of X-ray emission.

\subsubsection{Accretion changes the magnetic field structure?}

Numerical simulations suggest that
the pressure of the accreting material can distort the large-scale
stellar magnetic field considerably \citep[e.g.][]{Miller97, Romanova04} and
the magnetospheric transfer of material to the star can give rise to
instabilities of the magnetic fields around the inner disk edge
and cause reconnection events.
The presence of accreting material also leads to higher densities in
(parts of) the magnetosphere.
In contrast to the WTTS, which probably have loops with moderate
plasma densities,
some fraction of the magnetic field lines in CTTS would be mass loaded
and therefore have much $(\sim 100\times)$ higher densities.
If now a magnetic reconnection event liberates a certain amount of
energy, this can heat the plasma in the low-density loops of WTTS
to X-ray emitting temperatures ($\gtrsim 10$~MK), 
while the denser plasma in the mass loaded loops of the CTTS would be only
heated to much lower temperatures, and remain too cool to emit X-rays.
This effect may cause the lower X-ray luminosities of the CTTS as 
compared to WTTS.

However, a more quantitative assessment of this model is difficult.
According to magnetospheric accretion models, the fraction of the
stellar surface that is covered by accretion funnels should be 
at most a few percent \citep[e.g.,][]{Calvet98, Muzerolle01}.
This may be a too small fraction
to explain the reduction of the X-ray luminosity by a factor of
$\sim 2$. On the other hand, we note that the estimates of the area of 
accretion funnels are uncertain, and  other factors like 
global changes in the topology of the magnetic field may play
a (more?) important role. It therefore  seems possible that magnetospheric 
accretion streams are
somehow related to the different X-ray activity levels of accreting
and non-accreting stars.

It is also interesting to note that the analysis of the
largest flares in the COUP data by Favata et al.~(2005) seems to indicate
that intense, active accretion may inhibit magnetic heating of the
accreting plasma, while in stars which are not actively accreting the
long magnetic structures may acquire a ``coronal'' character.

\subsubsection{Accretion changes the stellar structure?}

The accretion process may alter the internal stellar structure and 
the differential rotation patterns, and thereby influence the 
magnetic field generation process.
For example, \cite{Siess99} found in their stellar evolution calculations 
that accretion reduces the efficiency of convection.
This theoretical result agrees with another finding 
based on comparison of orbital masses of PMS stars
with  evolution models by \cite{Stassun04b}, who found
 that TTS seem to have lower convection efficiencies
than MS stars. The reduced convection efficiency may
lead to weaker dynamo action.

Another effect may be that 
the magnetospheric star-disk coupling affects the 
differential rotation pattern 
at the stellar surface. If the coupling happens closer to 
the stellar equator than to the poles, the magnetospheric braking effect 
thought to be at work in CTTS
may reduce the amount of differential rotation, and this also may decrease
the efficiency of the dynamo. 

Finally, the magnetic star-disk interaction may also have an effect on
the coupling between inner and outer
layers within the star, and thereby affect the level of magnetic activity
\citep[see][]{Barnes03a, Barnes03b}.

\subsection{Implications for pre-MS magnetic dynamos}

The MS activity-rotation relation is well-established
in stars \citep[e.g.,][]{Pallavicini81, Pizzolato03} and is usually interpreted in terms of the
$\alpha\!-\!\Omega$-type dynamo that is thought to work in the Sun.
A simplified description of the rather complicated processes by which 
this dynamo generates surface magnetic fields 
can be summarized as follows (for details see, e.g., \citealp{Schrijver00, Ossendrijver03}):
The strong differential rotation in the tachocline, a region near the 
bottom of the convection zone in which the rotation
rate changes from being almost uniform in the radiative interior
to being latitude dependent in the convection zone,
generates strong toroidal magnetic fields.
While most of the toroidal magnetic flux is stored and further 
amplified in the tachocline, 
instabilities expel individual flux tubes, which then 
rise through the convection zone, driven by magnetic buoyancy, until 
they emerge at the surface as active regions.
The power of the dynamo (i.e.~the magnetic energy created 
by the dynamo per unit time) depends only indirectly
on the rotation rate.  The $\alpha-\Omega$ dynamo is
principally dependent on the
radial gradient of the angular velocity in the
tachocline and the characteristic scale length of convection at the base 
of the convection zone.
The empirical relationship between X-ray luminosity and 
rotation rate in MS stars does therefore {\em not}
mean that the power of the dynamo scales with rotation rate,
but rather
that faster rotating stars have stronger velocity shear in the
thin tachoclinal layer between the radiative core and the outer convective
zone. 

The presence of an $\alpha\!-\!\Omega$-type dynamo
at the bottom of the convection zone does not prevent 
other dynamo processes from {\em also} operating in a star. 
In the Sun,
small scale turbulent dynamo action \citep[e.g.][]{Durney93}
 is taking place throughout the convection
zone and is thought to be responsible for the small-scale intra-network fields.
Recent results \citep{Bueno04} suggest that the total magnetic flux 
generated by the 
small-scale turbulent dynamo action is much larger than previously assumed.
This means that two conceptually distinct magnetic dynamos 
are simultaneously operating in the contemporary Sun. 
In the case of the Sun, the
coronal activity is most likely dominated by the tachoclinal dynamo action.
Most of the ONC TTS, however, are thought to be
fully convective, or nearly fully convective,
so the tachoclinal layer is either buried very deeply, or does not
exist at all.  
It is therefore reasonable to assume that in these
(nearly) fully convective TTS, a convective dynamo is the
main source of the magnetic activity.

An interesting possible alternative explanation may be that the 
conventional wisdom, i.e.~that TTS are fully convective, is not correct.
We note that several studies have shown that accretion can significantly 
change the stellar structure. For example, \cite{Prialnik85}
found that even for moderate accretion rates
the stars are no longer fully convective.
More recently, \cite{Wuchterl03} 
found that accreting PMS stars are not fully convective;
their model of a solar mass star at 1 Myr has a radiative core
and a convective envelope, resembling the present Sun rather than
a fully convective object.
These results open the possibility for a solar-type tachoclinal dynamo
to work in the TTS. Our results on the relation between X-ray activity
and Rossby number are not inconsistent with that possibility.
\bigskip

We provide
in this study various empirical relationships of the X-ray
luminosities with, e.g., stellar mass, or bolometric luminosity,
which should also be relevant to the dynamos operating in TTS.
A purely empirical explanation of these
correlations is given by the existence of 
upper and lower limits to the X-ray activity levels, in analogy
to results for MS stars. The upper limit is caused by the
saturation level of magnetic activity around 
$\log\left(L_{\rm X}/L_{\rm bol}\right) \sim -3$
\citep[e.g.,][]{Pizzolato03}.
A lower limit is suggested by studies of 
nearby field stars, that led to the conclusion that
all cool dwarf stars are surrounded by X-ray emitting coronae
with a minimum X-ray surface flux of about 
$1\times 10^4\,{\rm erg/cm^{2}/sec}$ \citep{Schmitt97, Schmitt04};
for early M-type TTS in the ONC this surface flux corresponds to 
$\log\left(L_{\rm X}/L_{\rm bol}\right) \sim -6$.
The restriction of $\log\left(L_{\rm X}/L_{\rm bol}\right)$ to the range
between about $-6$ and  $-3$ leads to correlation between X-ray luminosity and
bolometric luminosity;
the correlation between X-ray luminosity and stellar mass
can then be explained by the dependence of bolometric luminosity on stellar mass.
An alternative explanation for the correlations can be based on the
finding that the fractional X-ray luminosities
increase with stellar mass (\S\ref{lx_mass.sec}).
This is consistent with the results of \cite{Pizzolato03}, who showed
that in low-mass MS stars the saturation level
in $L_{\rm X}/L_{\rm bol}$ increases with stellar mass.
These results suggest a similar origin of X-ray activity in the TTS and
MS stars, and thus provide support
for the standard $\alpha-\Omega$ solar-type-dynamo model for
TTS X-ray emission.

\section{Summary}

The main results from our study of the X-ray properties of the TTS
in the ONC can be summarized as follows:

In the COUP data we detect X-ray emission from essentially every late-type
(F to M) ONC star.  There is no indication for the existence of an ``X-ray quiet'' 
population of stars with suppressed magnetic activity.
We find that the X-ray luminosities of the TTS are correlated to 
bolometric luminosities,
stellar masses, and effective temperatures.
The $L_{\rm X}\leftrightarrow M$ correlation for the
TTS shows a slope similar to the corresponding correlation
for MS stars, which is
probably related to the association between mass and MS
X-ray saturation levels. 
Together, these lines of evidence
suggest that the $L_{\rm X}\leftrightarrow M$ relationship may be more
physically fundamental than X-ray relationships to
bolometric luminosity, surface area, or rotation.

Our data indicate a correlation between 
X-ray activity and rotation period, apparently in strong contrast
to the well established anti-correlation seen for MS stars.
However, the efficacy of our analysis is limited since
rotation periods are only known 
for about 40\% of the TTS in our sample, and the missing stars
(i.e.~those with unknown rotation periods) probably introduce a bias.
If we consider Rossby numbers, we find that all TTS are located
in the saturated or super-saturated regime of the 
activity$\leftrightarrow$Rossby number relation for MS stars. 
In principle, the TTS may thus follow the same relation between 
X-ray activity and Rossby number as MS stars,
but the large scatter in $L_{\rm X}/L_{\rm bol}$ at any given 
Rossby number suggests that other factors are also involved in 
determining the level of X-ray activity.

The enormous scatter we generally find in the correlations between X-ray 
activity and other stellar parameters is larger than what one would
expect due to X-ray variability, uncertainties in the variables,
and the effects of unresolved binaries.
Therefore, this wide scatter must be related
to intrinsic differences in the individual TTS, and we find here
that the influence of accretion on the X-ray emission seems to play
an important role.  There is a 
remarkable contrast between the X-ray properties of accreting and
non-accreting stars:
Our data confirm previous results that accreting stars are less X-ray active 
than non-accreting stars (although a statistically significant difference is 
only found for stars in the $\sim 0.2-0.5\,M_\odot$ mass range)
and suggest an anti-correlation between fractional X-ray luminosity
and accretion rate.
The non-accreting TTS have the same median
X-ray activity level as rapidly rotating MS stars and show good
$L_{\rm X}\leftrightarrow L_{\rm bol}$  and $L_{\rm X}\leftrightarrow M$ correlations
with a scatter as expected from the uncertainties, X-ray variability, and
unresolved binaries.
The accreting TTS, on the other hand, show about 3 times lower X-ray activity 
levels and produce much less well defined 
$L_{\rm X}\leftrightarrow L_{\rm bol}$  and $L_{\rm X}\leftrightarrow M$ 
correlations with much wider scatter.
These findings imply that the apparent X-ray deficit of the whole TTS sample
(i.e.~the median fractional X-ray luminosity of
$\log\left(L_{\rm X}/L_{\rm bol}\right) \sim -3.5$, which is
below the
saturation limit around $\log\left(L_{\rm X}/L_{\rm bol}\right) \sim -3.0$
typically found for rapidly rotating MS stars)
is solely due to the reduced X-ray activity of the accreting TTS.

We discuss several possible explanations for the
suppression of X-ray emission in accreting stars.  The
effect may be related to changes of the coronal structure
or the internal stellar structure induced by the accretion
process.  We favor
the idea that magnetic reconnection can not heat the dense
plasma in mass-loaded accreting field lines to X-ray
temperatures.

The geometry of X-ray producing magnetic fields is
also still uncertain.  Solar-type coronal loops
are probably the dominant source of the observed X-ray emission.
However, we note that the study of the most powerful 
flares seen in COUP stars (Favata et al.~2005) suggests that in some objects
star-disk field lines extending
$>10 \times R_\ast$ from the stellar surface may be involved. 
 Accretion shocks at
the stellar surface can not be responsible for the
emission seen in COUP sources.
    Finally, the ultimate origin 
of the X-ray activity of the TTS is most likely either 
a turbulent dynamo working in the stellar convection zone, or, 
if theoretical suggestions that accreting TTS may not be fully convective
are correct, a solar-like $\alpha-\Omega$ dynamo at the base of the
convection zone.

\acknowledgements
COUP is supported by $Chandra$ Guest Observer grant SAO GO3-4009A (E.
Feigelson, PI). Further support was provided by
the Chandra ACIS Team contract NAS8-38252.
We would like to thank L.A.~Hillenbrand for useful comments on the manuscript,
J.H.M.M.~Schmitt
and C.~Liefke for information on the NEXXUS database, H.~Peter
for enlightening discussion about the solar corona, M.~H\"unsch
for information about the X-ray luminosities of subgiants,
and H.~Shang for discussions about accretion processes.
YCK has been supported by Korean
Research Foundation Grant KRF-2002-070-C00045.
BS, EF, GM and SS acknowledge financial support from an italian MIUR
PRIN program and an INAF program for the years 2002-2004.

Facility: CXO(ACIS)

\bibliography{aj-jour}

\newpage

\begin{deluxetable}{ccccccccc}
\centering \tabletypesize{\small} \tablewidth{0pt}
\tablecolumns{9}

\tablecaption{Detection fractions of ONC stars in the optical sample\label{detfrac.tab}}

\tablehead{

\colhead{Spectral type} & \colhead{O} & \colhead{B} &
\colhead{A} & \colhead{F} & \colhead{G} & \colhead{K} & \colhead{M0--6.5} 
& \colhead{all} }

\startdata
detections &  2  & 11     &3     &1     &8   &133  & 441      &  598  \\
non-detections&  0  &  1     &3     &0     &0   &  8  &  29      &   41  \\ 
${\cal F}$& 100\%& 91.7\%& 50.0\%&100\%&100.0\%&94.3\%&93.6\% & \\
confusion & 0   & 0     &2     &0     &0    & 6   & 17 &  24 \\
${\cal F'}$&  
           100\%& 91.7\%&75\%&100\%&100\% &98.5\%& 97.3\% \\ \hline
\multicolumn{9}{c}{In the lightly absorbed optical sample:} \\
detections &  2  & 11     &2     &1     &7   &117  & 414      &  554 \\
non-detections&  0  &  1     &3     &0     &0   &  3  &  26      &   32  \\
${\cal F}$ & 100\%& 91.7\%& 40.0\%&100\%&100\%&97.5\%&94.1\% & 94.5\%\\
confusion & 0   & 0     &2     &0     &0    & 3   & 15 &  19 \\
${\cal F'}$ &  
           100\%& 91.7\%&66.7\%&100\%&100\% &100\%& 97.4\% & 97.7\%\\

\enddata

\tablecomments{${\cal F}$ is the detection fraction.
${\cal F'}$ is the detection fraction if objects with non-detections due to
X-ray confusion are removed from the sample, which gives the fraction
of objects that are below the X-ray detection limit. 
The COUP undetected K6-star H97-9320, which lies far (2.7 mag) below the 
ZAMS in the HR-diagram, has been removed from the optical sample.
 }

\end{deluxetable}

\pagebreak

\begin{figure}[p]
\plotone{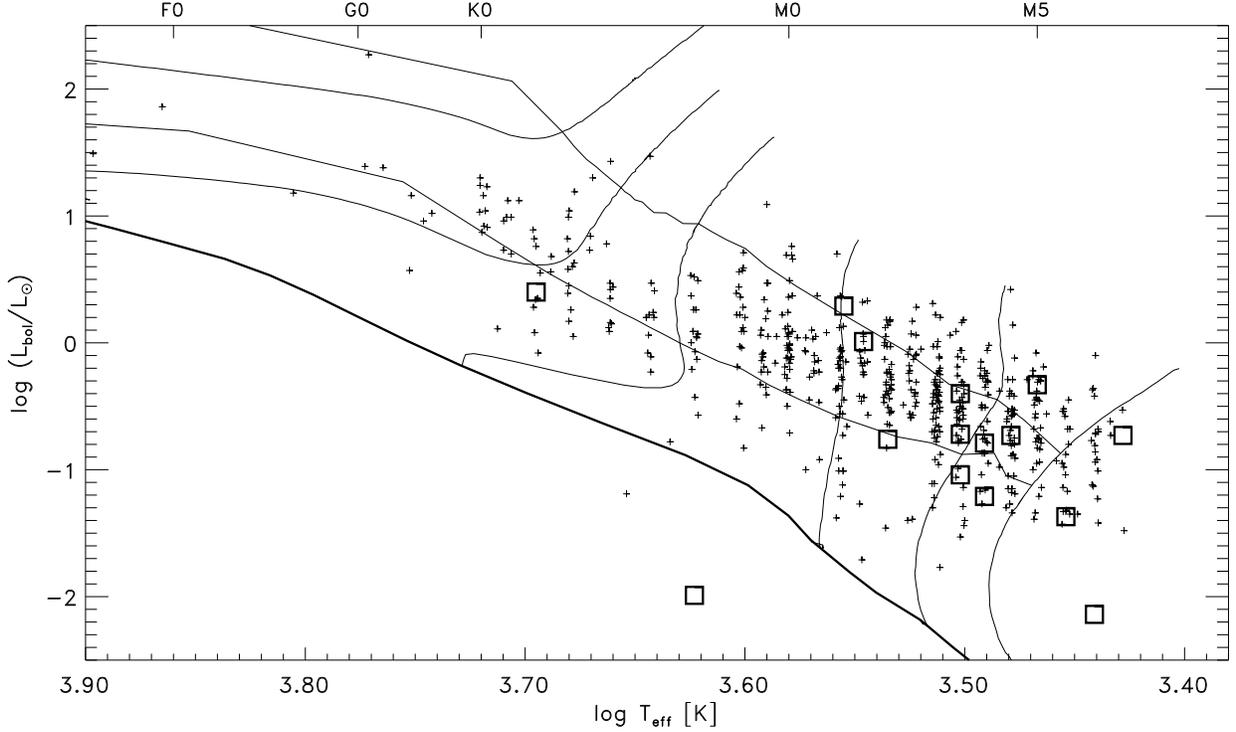}
\figcaption{HR-diagram for the late type stars in the optical sample.
X-ray detected objects
are plotted as crosses. To avoid the strong overlap of objects with the
same spectral types, the $\log\left(T_{\rm eff}\right)$ values
have been shifted by random numbers in the range $[-0.002\,\dots\,
+0.002]$.
Members of the optical sample which are
not detected as X-ray sources and not affected by X-ray source confusion
are marked by open squares. 
The lines show isochrones for ages
of $3 \times 10^{5}$~years and  $3 \times 10^{6}$~years
and the ZAMS, and PMS tracks for stellar masses of $0.1, 0.2, 0.4, 1, 2, 4\,M_\odot$
according to the evolutionary models of \citet{Siess00}. Note that the
X-ray undetected late K-type star well below the ZAMS is very likely not
a member of the ONC.
\label{hrd.fig}}
\end{figure}
\clearpage

\clearpage

\clearpage

\clearpage

\clearpage

\clearpage

\clearpage

\clearpage

\clearpage

\clearpage

\clearpage

\clearpage

\clearpage

\clearpage

\clearpage

\clearpage

\clearpage

\end{document}